\documentclass[letterpaper,twocolumn,10pt]{article}

\usepackage{usenix, times,color,graphicx,amsmath,array,booktabs, amssymb, subcaption}
\usepackage[normalem]{ulem}
\usepackage[
  backend=biber,
  doi=false,
  eprint=false,
  natbib=true,
  style=trad-abbrv
]{biblatex}
\usepackage{tikz}
\usepackage{varioref}
\usepackage{colortbl}
\usepackage[normalem]{ulem}
\usepackage{xcolor}
\usepackage{multirow}
\usepackage{comment}
\usepackage{hyperref}
\usepackage{algorithmicx}
\usepackage[ruled, vlined, linesnumbered]{algorithm2e}
\usepackage{url}
\usepackage{amssymb}

\usepackage{amsmath}
\usepackage{wrapfig}
\usepackage{amsfonts}
\usepackage{amsthm}
\usepackage{listings}
\usepackage{xspace}
\usepackage{tightenum}
\usepackage{adjustbox}
\usepackage{listings}
\usepackage{microtype}
\usepackage{tabulary}
\usepackage{floatflt}
\usepackage{mdframed} 
\usepackage{textcomp}
\usepackage[small,compact]{titlesec}
\usepackage{enumitem}
\usepackage[normalem]{ulem}
\usepackage[T1]{fontenc}
\usepackage[utf8]{inputenc}
\usepackage{tikz}
\usepackage{mdframed}
\usepackage{layouts}

\usetikzlibrary{
  arrows,
  calc,
  matrix,
  positioning,
  shapes,
}

\hypersetup{
  colorlinks,
  linkcolor={red!50!black},
  citecolor={red},
  urlcolor={blue!80!black}
}

\newcommand\blfootnote[1]{%
  \begingroup
  \renewcommand\thefootnote{}\footnote{#1}%
  \addtocounter{footnote}{-1}%
  \endgroup
}

\newcommand{\system}[0]{Bamboo\xspace}
\newcommand{\spotinstances}[0]{spot instances}

\newcommand{\codeIn}[1]{\texttt{#1}}

\newcommand{\figref}[1]{Figure~\vref{fig:#1}}

\newcommand{\hx}[1]{{\color{red} HX: #1}}

\newcommand{\eg}{\hbox{\emph{e.g.}}\xspace}
\newcommand{\ie}{\hbox{\emph{i.e.}}\xspace}

\newcommand{\MyPara}[1]{\vspace{.1em}\noindent\textbf{\textit{#1}}~~}
\begin{document}






\title{\system{}: Making Preemptible Instances Resilient for Affordable Training of Large DNNs}
\author{John Thorpe$^{\dag\clubsuit}$\hspace{1.4em}Pengzhan Zhao$^{\dag\clubsuit}$\hspace{1.4em}Jonathan Eyolfson$^{\dag}$\hspace{1.4em}Yifan Qiao$^{\dag}$\hspace{1.4em}Zhihao Jia$^{\ddag}$\\[.3em]Minjia Zhang$^{\S}$\hspace{2.5em}\rm{Ravi Netravali}$^{\ast}$\hspace{2.5em}Guoqing Harry Xu$^{\dag}$
\\[.3em]
UCLA$^{\dag}$\hspace{1.5em}CMU$^{\ddag}$\hspace{1.5em}Microsoft Research$^{\S}$\hspace{1.5em}Princeton University$^{\ast}$ \\[1em]
}
\maketitle

\pagestyle{plain}
\pagenumbering{arabic}


\begin{abstract}

DNN models across many domains continue to grow in size, resulting in high resource requirements for effective training, and unpalatable (and often unaffordable) costs for organizations and research labs across scales. This paper aims to significantly reduce training costs with effective use of preemptible instances, i.e., those that can be obtained at a much cheaper price while idle, but may be preempted whenever requested by priority users. Doing so, however, requires new forms of \emph{resiliency} and \emph{efficiency} to cope with the possibility of frequent preemptions --  a failure model that is drastically different from the occasional failures in normal cluster settings that existing checkpointing techniques target.

We present \system, a distributed system that tackles these challenges by introducing \emph{redundant computations} into the training pipeline, \ie, whereby one node performs computations over not only its own layers but also over some layers in its neighbor. Our key insight is that training large models often requires \emph{pipeline parallelism} where ``pipeline bubbles'' naturally exist. \system{} carefully fills redundant computations into these bubbles, providing resilience at a low cost. Across a variety of widely used DNN models, \system{} outperforms traditional checkpointing by \textbf{3.7$\times$} in training throughput, and reduces costs by \textbf{2.4$\times$} compared to a setting where on-demand instances are used.
\end{abstract} 
\blfootnote{$^\clubsuit$ Contributed equally.}
\section{Introduction}
\label{sec:introduction}

DNNs are becoming progressively larger to deliver improved predictive performance across a variety of tasks, including computer vision and natural language processing.
For instance, recent language models such as BERT
\cite{corr/abs-1908-08962} and GPT \cite{radford2019language} already have a massive number of parameters, and their newer variants continue to grow at a rapid pace. For example,
BERT-large has 340 million parameters, GPT-2 has 1.5 billion, and GPT-3
increases to 175 billion; the next generation of models embed upwards of trillions of parameters
\cite{corr/abs-2101-03961}.

Of course, model growth also entails larger training costs. For instance, GPT-3 consumes several thousand petaflop/s-days, costing over \$12
million to train on a public cloud (needing hundreds of GPU servers)
\cite{nips/2020/brown}. Unfortunately, such costs are prohibitive for small organizations. Even for large tech firms, training today's models incurs an exceedingly high monetary cost that eventually gets billed to the training department. While pretrained models may be reused and fine-tuned for different applications, training new models is often required to keep pace with changing or emerging workloads and datasets.

Although there exists a body of work on improving the training of large models~\cite{pipedream-sosp19,narayanan-osdi20,checkmate-mlsys20,chetlur2014cudnn,all-reduce-16, dawnbench-sigops19,geeps-eurosys16,minibatch-sgd-17,seide2014-bit-interspeech14,speech-dnns-icassp14,communication-mpich-jhpca05,poseidon-atc17,gpipe-18,flexflow-mlsys18, parallelize-cnn14}, existing techniques focus primarily on \emph{scalability} and \emph{efficiency}, with monetary costs often being neglected.  However, when \emph{affordability} and \emph{accessibility} are considered, resource usage becomes a key concern and none of these techniques were targeted at improving \emph{cost-efficiency} (\eg, performance-per-dollar) for training.

\MyPara{Preemptible Instances.} This paper explores the possibility of using preemptible instances\textemdash a popular class of cheap cloud resources\textemdash to reduce the cost of training large models. There are several kinds of preemptible instances. For example, major public clouds  provide \emph{spot instances} with a price much cheaper than \emph{on-demand instances}\textemdash \eg, the hourly rate of a GPU-based spot instance is only $\thicksim$30\% of that for its on-demand counterpart on Amazon EC2~\cite{spot-price}. As another example, large datacenters often maintain certain amounts of compute resources that can be allocated for any non-urgent tasks but will be preempted as urgent tasks arise~\cite{ras-sosp21,boucher-atc20}. Similarly, recent ML systems~\cite{jeon-atc19, antman-osdi20, pipeswitch-osdi20} allow training jobs to use inference-dedicated machines to fully utilize GPU resources but preempts those machines when high-priority inference jobs arrive. The presentation of this paper focuses on spot instances, but we note that our techniques are generally applicable to any type of preemptible resources.

Despite their substantial cost benefits, preemptible instances pose major challenges in reliability and efficiency due the frequent and unpredictable nature of their preemptions. When and how many instances get preempted depends primarily on the number of priority jobs/users in a cluster. 
In a public spot market, preemption can also result from the market price exceeding the user's bid price.
While price-based preemption can be avoided via a high bid price (\eg, the on-demand price), capacity-based preemption is unavoidable. Preemption patterns vary drastically across clouds and even across families/zones on the same cloud (\S\ref{sec:motivation}). 


Given the unpredictable nature of spot instances, users can often only run short, stateless jobs and simply restart these jobs if they get preempted. Model training, on the contrary, is stateful and time-consuming. Discarding the state (\eg, learned weights) upon each instance preemption not only wastes computation but also prevents training from making progress. Checkpointing-based techniques can reduce wasted computation to a degree, but still spend a significant fraction of the training time (\eg, \textbf{77\%} when training GPT-2 with 64 EC2 spot instances, see \S\ref{sec:motivation}) on restarting and redoing prior work in the presence of frequent preemptions~\cite{tributary-atc18, proteus-eurosys17}\textemdash a largely different scenario compared to conventional clusters where failures are rare.



\MyPara{\system.} This paper presents \system, a distributed system that provides resilience and efficiency for DNN training over preemptible instances. 
\system supports both pipeline parallelism and (pure) data parallelism with the same approach. Since pipeline parallelism is a more complex and general approach (for training large models), our discussion focuses on pipeline parallelism; we briefly discuss our support for pure data parallelism in \S\ref{sec:data}. \system currently does \emph{not} support model parallelism. 

\MyPara{Redundant Computation.} Key to the success of \system is a set of novel techniques centered around \emph{redundant computation} (RC), inspired by how disk redundancies such as RAID~\cite{raid} provide resilience in the presence of disk failures. A training system that uses pipeline parallelism runs a set of data-parallel pipelines, each training on a partition of the dataset. Each node\footnote{In this paper, ``instance'' and ``node'' both refer to a spot instance.} in a data-parallel pipeline performs (forward and backward) computations over a shard of NN layers with a \emph{microbatch} of data items~\cite{gpipe-18}. 
\system lets each node in each data-parallel pipeline carry its own shard of layers as well as its successor's shard. Each node performs \emph{normal} computation over its own layers and \emph{redundant} computation over its successor's layers. The reason why we use a neighbor node  (as opposed to a random node) to run RC is to exploit data locality in pipeline parallelism (see \S\ref{sec:redundant-computation-and-recovery}).
Upon a node preemption, its predecessor has all the information (\eg, layers, activations) needed for the training to progress; continuing training requires running a failover schedule on the predecessor node without wasting prior computations. 

At first glance, running RC on every node appears infeasible due to concerns with both time and memory. \system overcomes these challenges by taking into account pipeline characteristics to carefully reduce/hide these overheads.

\begin{figure*}[t]
    \centering
    \begin{tabular}{ccc}
         
    \includegraphics[scale=1]{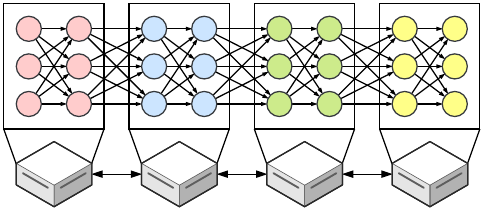} &
     \includegraphics[scale=1]{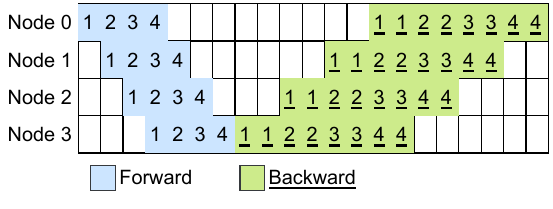} &
     \includegraphics[scale=1]{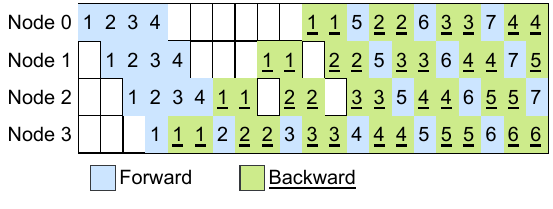}\\
     (a) Pipeline parallelism & (b) GPipe scheduling & (c) PipeDream scheduling
    \end{tabular}
    \caption{Illustration of pipeline parallelism on a 4-node cluster: (a) the model is divided into 4 shards, each with 2 layers; (b) and (c) show the scheduling of two recent systems GPipe~\cite{gpipe-18} and PipeDream~\cite{pipedream-sosp19}. \label{fig:pipeline}}
\end{figure*}

First, to minimize the time overhead from RC, \system leverages a key insight that \emph{bubbles}~\cite{gpipe-18, deepspeed-sc20} inherently exist in systems using synchronous pipeline parallelism (\S\ref{sec:background}).
Bubbles are idle times on each node due to the gaps between the forward and backward processing of microbatches (Figure~\ref{fig:pipeline}).  \system schedules the forward redundant computation (FRC) on each node asynchronously into the bubble. For the part of FRC that cannot fit into the bubble, \system overlaps it with the normal computation. As a result, FRC incurs a tolerable overhead (\ie, no extra communication is needed due to locality, and it can overlap with normal computation), and hence \system performs it \emph{eagerly} in each epoch. If a node is preempted during a forward pass, the pipeline continues after a node rerouting step whose overhead is negligible. 

Unfortunately, for backward redundant computation (BRC), such a bubble does not exist. Eager BRC would require much extra work and data-dense communication on the critical path, which could delay training significantly (\S\ref{sec:redundant-computation-and-recovery}).  
As such, \system runs BRC \emph{lazily} only when a preemption actually occurs. If a node is preempted in a backward pass, continuing the pipeline requires a pause for the node's predecessor to perform BRC to restore the lost state. However, since FRC is performed eagerly, when BRC runs, much of what it needs is already in memory, keeping pauses short.


Second, performing RC increases each node's GPU memory usage. Note that the major source of the memory overhead is storing intermediate results (activations and optimizer state) from FRC, \emph{not} the redundant layers, which take only little extra memory. 
To mitigate the memory issue, we leverage \system's unique way of performing RC described above. Note that the purpose of saving intermediate results of a forward pass is that these results are used by its backward computation. However, in \system, BRC is performed lazily upon preemptions and the intermediate results of FRC are thus not needed in normal backward passes.  Hence, 
\system swaps out the intermediate results of each node's FRC into the node's CPU memory, leading to substantial reduction in GPU memory usage. These results are swapped back into GPU memory for BRC only upon preemptions. 




\system continues normal training with the help of RC in the presence of non-consecutive preemptions, \ie, preempted instances are not neighbors in the same data-parallel pipeline. Once consecutive instances are preempted, RC can no longer provide resilience. More redundancies could be added to provide stronger resilience, but this would incur (compute and communication) overheads that are too significant to hide. Instead, based on our empirical observation that most concurrent preemptions come from the same allocation group (\eg, a zone), \system takes care to ensure that consecutive nodes in each pipeline come from different zones, minimizing the chance of consecutive preemptions at a small ($<$5\%) overhead (see \S\ref{sec:cross}). 

\MyPara{Reconfiguration.}  In cases where consecutive preemptions do happen, we must reconfigure the pipelines (\S\ref{sec:reconfiguration-and-transfer}). Further, even if preemptions are non-consecutive, continuing training with RC after many preemptions is a ``spare-tire'' approach, which is vulnerable to future preemptions. To solve these problems, \system provides a Kubernetes-based framework that monitors preemptions and reconfigures the pipelines by dynamically adding instances and adjusting pipeline configurations (\eg, the number of pipelines). 
\system checkpoints the model state periodically. If no allocations can be made (\ie, a rare situation where the cluster is exhausted) and the remaining nodes are too few to sustain the training, \system suspends the training until enough new instances can be obtained and the training can restart from the checkpoint. 

\MyPara{Results.} We built \system atop DeepSpeed~\cite{deepspeed-sc20} and evaluated it by training 6 representative DNN models using EC2 spot clusters comprised of p3 instances. Compared to a baseline using on-demand instances, \system delivers a \textbf{3.6$\times$} cost reduction. \system also outperforms a checkpointing approach by \textbf{3.7$\times$}. We developed a simulation framework that takes preemption traces from real spot clusters and training parameters to simulate how training progresses with larger numbers of nodes. A deep-dive with BERT across a wide range of preemption probabilities shows that the \emph{value} (\ie, performance-per-dollar) \system provides stays constant and is much higher (\textbf{2.48$\times$}) than that of on-demand instances. 
\system will be open-sourced.

\section{Background\label{sec:background}}
This section discusses necessary background for parallelism strategies.
{\em Data parallelism} keeps a replica of an entire DNN on each device, which processes a subset of training samples and iteratively synchronizes model parameters with other devices.
Data parallelism is often combined with pipeline and/or model parallelism to train large models that do not fit on a single device.
{\em Model parallelism}~\cite{DistBelief} partitions model operators across training devices. However, efficient model parallelism algorithms are extremely hard to design, requiring difficult choices among scaling capacity, flexibility, and training efficiency.
As such, model-parallel algorithms are often architecture- and task-specific.

{\em Pipeline parallelism}~\cite{pipedream-sosp19, gpipe-18, pipemare-mlsys19} has gained much traction recently due to its flexibility and applicability to a variety of neural networks.
Pipeline parallelism divides a model at the granularity of layers  and assigns a shard of layers to each device. Figure~\ref{fig:pipeline}(a) shows an example where the model is partitioned into four shards and each worker hosts one shard (with two layers). 
Each worker defines a computation stage and the number of stages is referred to as the \emph{pipeline depth} (\eg, 4 in the example). One worker only communicates with nodes holding its previous stage or next stage.
Each input batch is further divided into {\em microbatches}. In each iteration, each microbatch goes through all stages in a forward pass and then returns in an opposite direction in a backward pass. There are often multiple microbatches residing in the pipeline and different nodes can process different microbatches in parallel to improve utilization. 


A key challenge in efficient pipeline parallelism is how to schedule microbatches. GPipe~\cite{gpipe-18} schedules forward passes of all microbatches before any backward pass, as shown in Figure~\ref{fig:pipeline}(b) where each node processes four microbatches. This approach leaves a "bubble" (\ie, white cells) in the middle of the pipeline, leading to inefficient use of compute devices. PipeDream~\cite{pipedream-sosp19} proposes the one-forward-one-backward (1F1B) schedule to interleave the backward and forward passes, as shown in Figure~\ref{fig:pipeline}(c). 1F1B can reduce the bubble size and the peak memory usage.

However, even with carefully-designed schedules, the pipeline bubble is still hard to eliminate. A fundamental reason is that it is extremely difficult to find the optimal layer partitioning to have each stage processed at the same rate. There exists a body of algorithms proposed recently to optimize layer partitioning and most of them are model- and hardware-specific \cite{pipedream-sosp19, dapple-ppopp21}. These algorithms are often time-consuming for large models, unsuitable for preemptible instances where the number of nodes keeps changing~\cite{varuna-2021}.



PipeDream~\cite{pipedream-sosp19} proposes asynchronous pipelining to eliminate the bubble\textemdash a node is allowed to work with stale weights to reduce the wait time. However, asynchronous microbatching introduces uncertainty in model convergence. In general, the effectiveness of synchronous v.s. asynchronous training is still open to debate. Furthermore, asynchronous training introduces inconsistencies in model state, which can create a more significant convergence issue when training occurs on preemptible instances, due to the need of frequent reconfigurations. For example, under synchronous microbatching, a reconfiguration can be performed at the end of each optimizer step (\ie, parameter update), and hence the reconfigured pipelines can start with the up-to-date parameters. This is impossible to do under asynchronous microbatching. 

As a result, we built \system{} atop synchronous microbatching where model state is always consistent. Instead of attempting to reduce the bubble, we explore an orthogonal direction\textemdash how to leverage the bubble to run RC efficiently.

\section{Motivation}
\label{sec:motivation}
This section motivates \system from two aspects: (1) high preemption rates and unpredictability of spot instances, and (2) high performance overheads of strawman approaches.


\begin{figure}[t]
  \centering
  \begin{tabular}{cc}
    \begin{minipage}[t]{.42\linewidth}
      \includegraphics[scale=0.25]{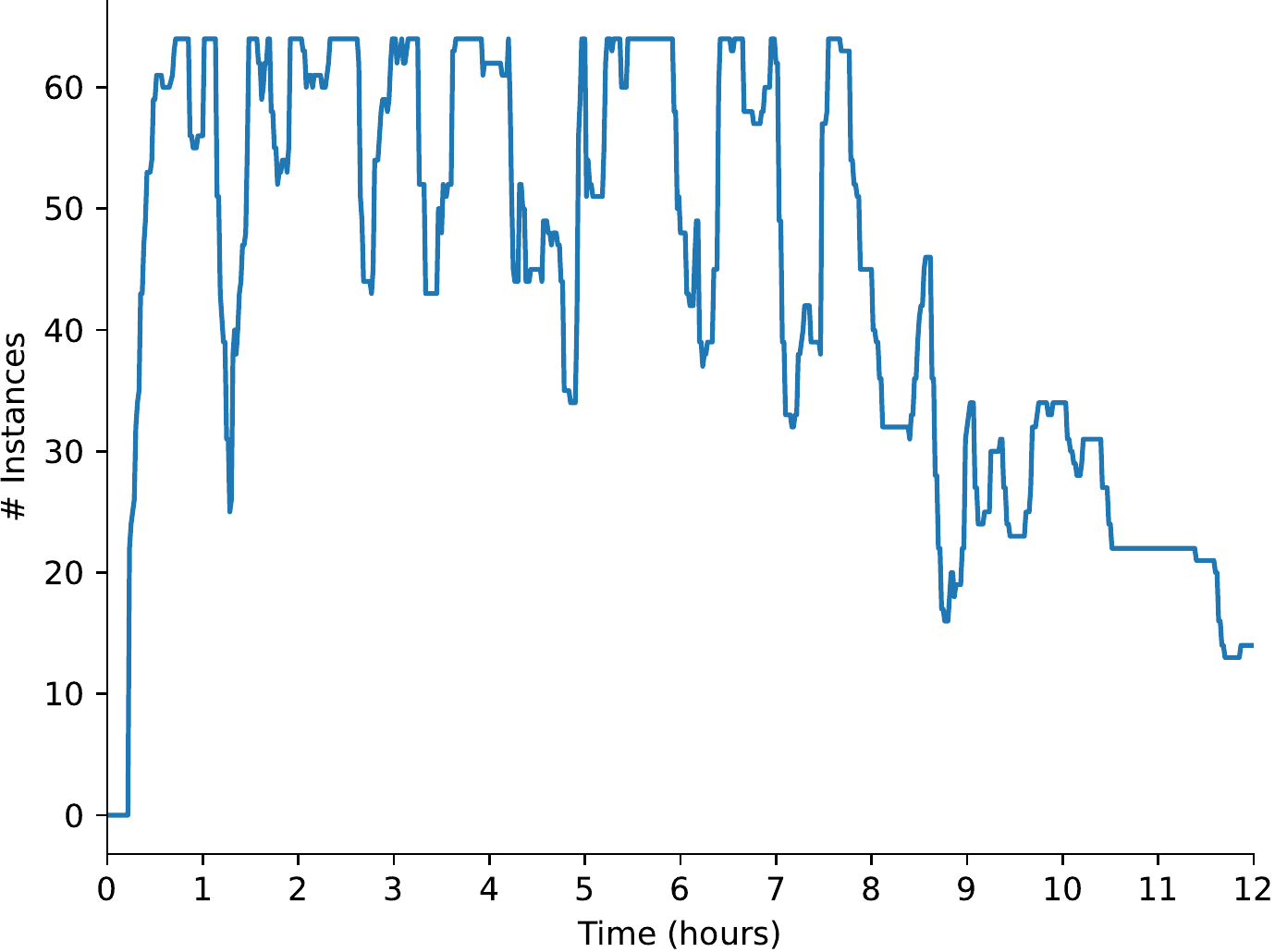}
    \end{minipage}
    &
    \begin{minipage}[t]{.42\linewidth}
      \includegraphics[scale=0.25]{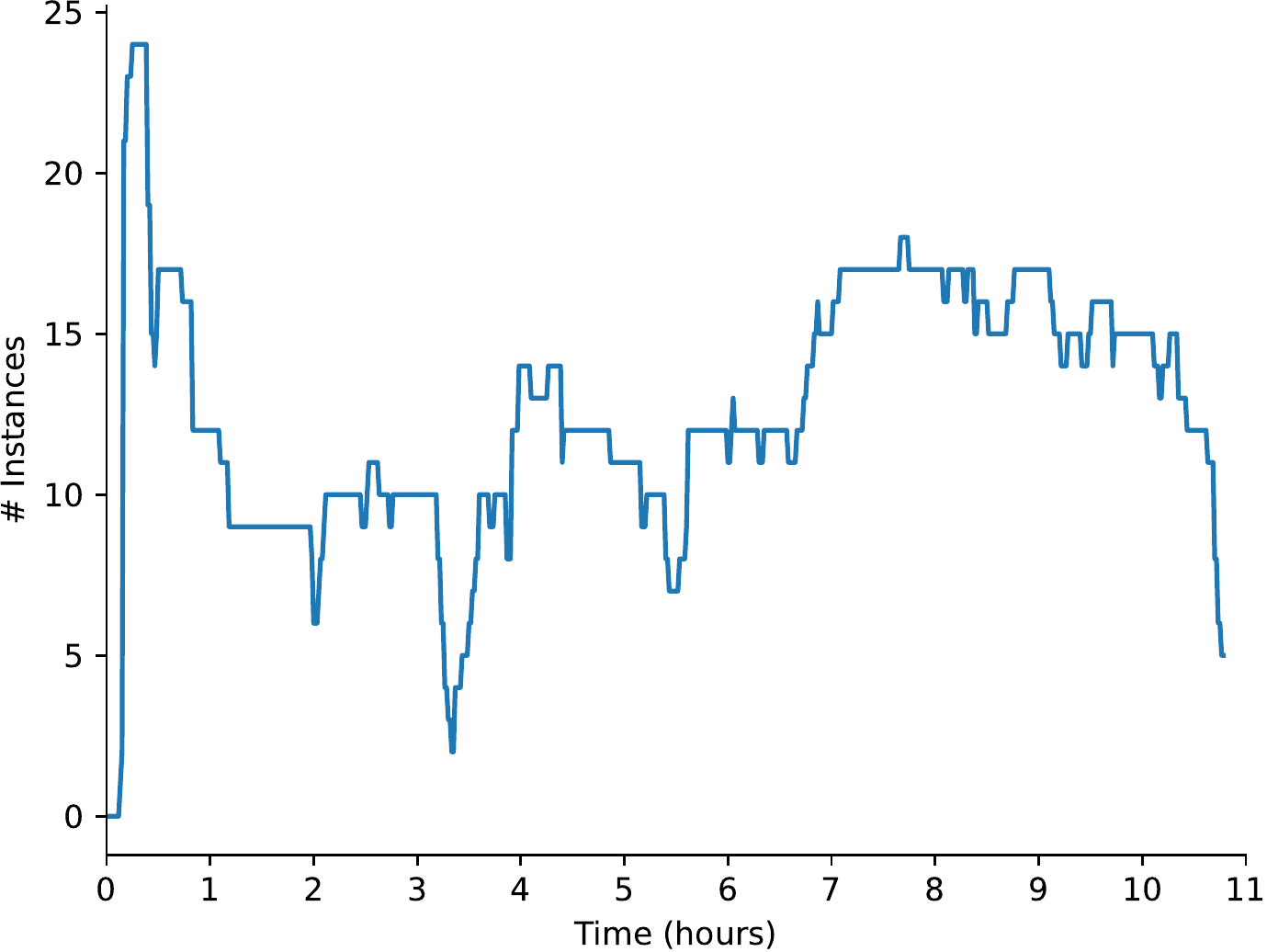}
    \end{minipage}
    \\
    (a) P3 @ EC2
    &
    (b) G4dn @ EC2
    \\
     \begin{minipage}[t]{.42\linewidth}
      \includegraphics[scale=0.25]{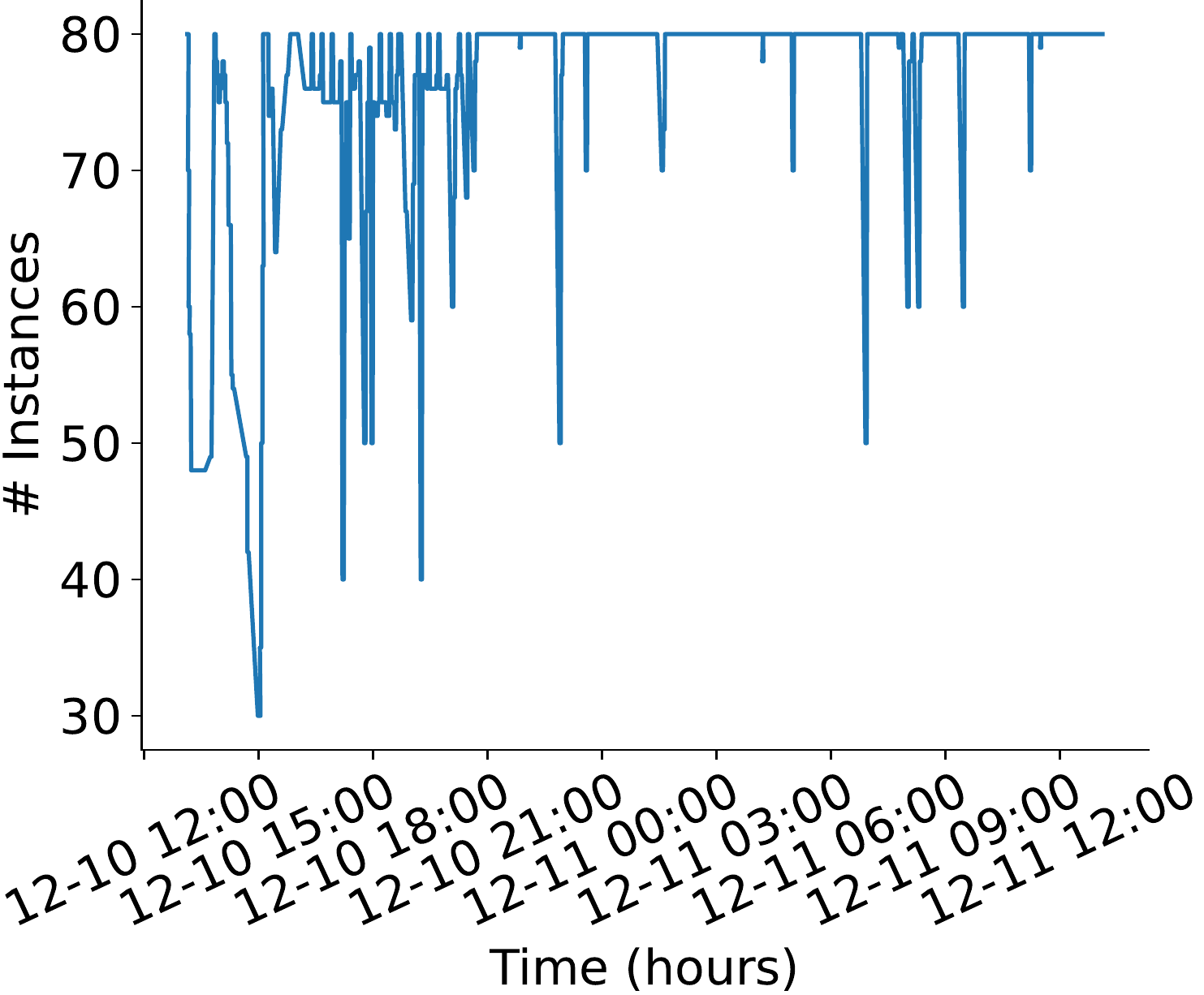}
    \end{minipage}
    &
    \begin{minipage}[t]{.42\linewidth}
      \includegraphics[scale=0.25]{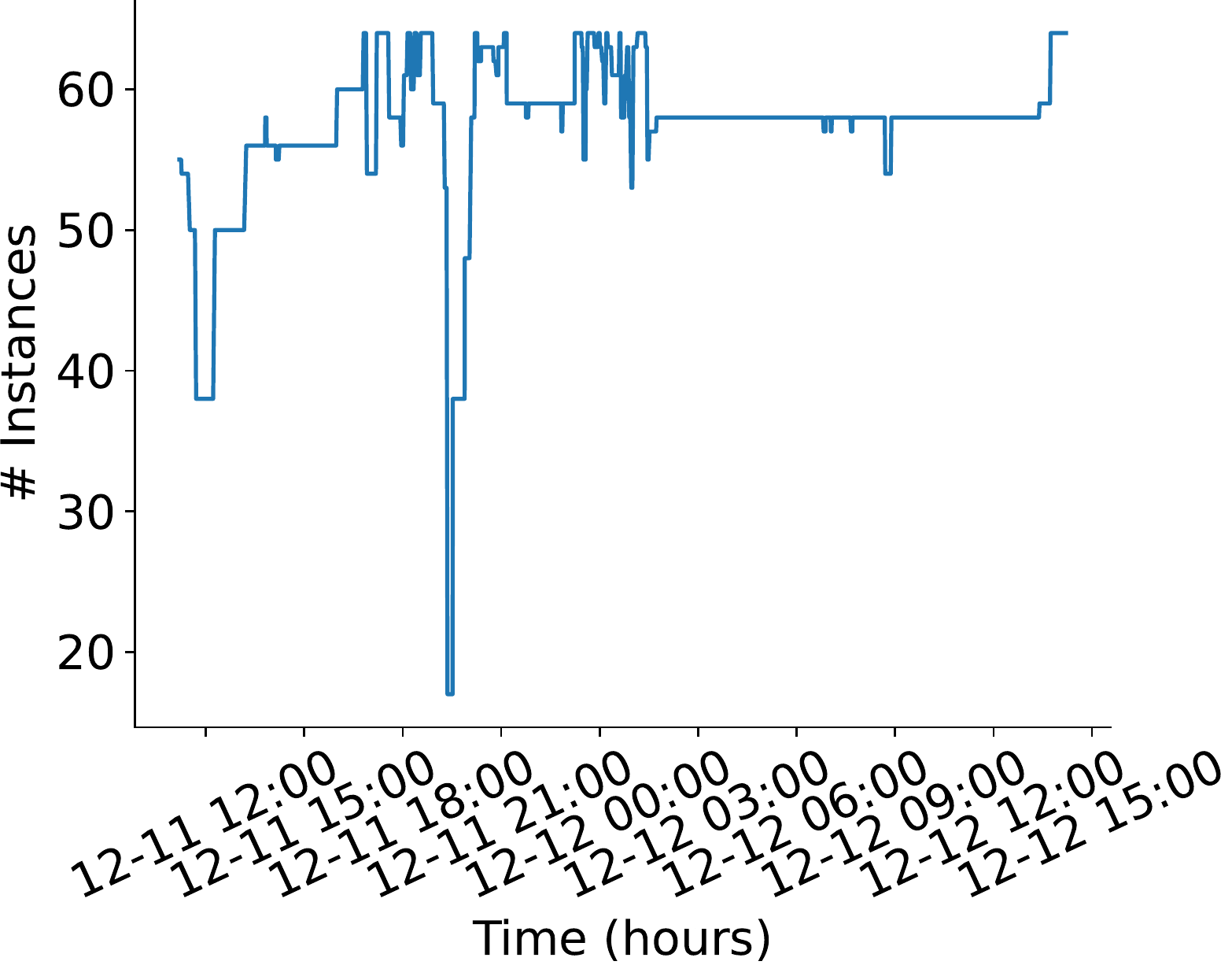}
    \end{minipage}
    \\
    (c) n1-standard-8 @ GCP & (d) a2-highgpu-1g @ GCP
    \\
  \end{tabular}
  \caption{
    Preemptions traces for a target cluster of size 64 instances on EC2 and 80 instances on GCP. Each graph shows a full-day trace for a GPU family in a cloud.
  \label{fig:preemptions_64}}
\end{figure}


\MyPara{Preemptions of Spot Instances.}
We first studied failure models with \spotinstances{} on
major public clouds.
Figure~\ref{fig:preemptions_64} shows a set of real preemption traces collected from running
spot instances in two public clouds: Amazon EC2 and Google Cloud Platform (GCP). For EC2, we used two GPU families: P3 (NVIDIA V100 GPUs with 32GB of memory) and G4dn (NVIDIA T4 GPUs with 16GB of memory). For GCP, we used \codeIn{n1-standard-8} (NVIDIA V100 GPUs with 16GB GRAM) 
and \codeIn{a2-highgpu-1g} (NVIDIA A100 GPUs with 40GB GRAM).
For each family, we collected traces for a 24-hour window. In each experiment, we used an autoscaling group to maintain a cluster of 64 with an exception of \codeIn{us-east1-c} in GCP, whose cluster size is 80. The autoscaling group, provided by each cloud, automatically allocates new instances upon preemptions to maintain the size (though without any guarantee). 

From both families, node
preemptions and additions are frequent and bulky (\ie, many nodes get preempted at each time). This can make a checkpointing-based approach restart many times in a short window of time, leading to large inefficiencies (discussed shortly). Furthermore, both preemptions and allocations are unpredictable. While the autoscaling group attempts to allocate new nodes to maintain the user-specified size, allocations are committed incrementally; new allocations are mixed with preemptions of existing instances, making the spot cluster an extremely dynamic environment. 

\begin{figure}[t]
  \centering
  \includegraphics[scale=0.5]{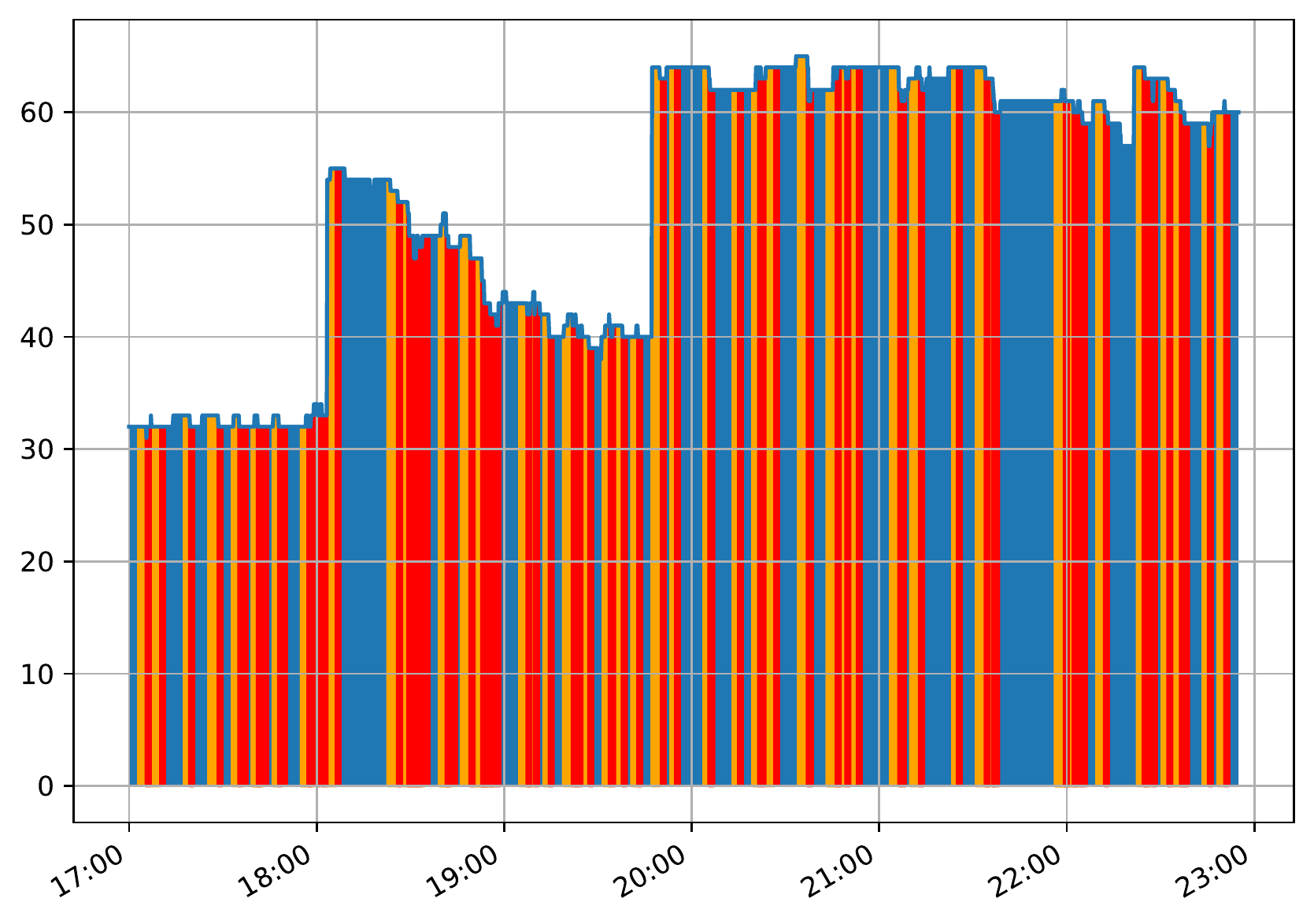}
  \caption{
    Training GPT-2 using checkpointing/restart with an
    autoscaling group of 64 P3 spot instances.
    Each color represents time spent in a distinct state,including
    \textcolor{blue}{Blue}: training actively made progress;
    \textcolor{orange}{Orange}: the cluster made progress that was then
    wasted; and
    \textcolor{red}{Red}: cluster restarting.
    \label{fig:checkpoint_perf}
  }
   \vspace{-1em}
\end{figure}

To understand the nature of the nodes that are preempted at the same time, we carefully analyzed two 24-hour preemption traces collected respectively from EC2 and GCP. For the EC2 trace, preemptions occur at 127 distinct timestamps, each of which see many preempted nodes. Of these 127 timestamps, only 7 see preemptions from multiple zones; at each of the remaining 120 timestamps, all nodes preempted come from the same zone. A similar observation was made on the GCP trace (12 out of 328 timestamps see cross-zone preemptions). These results confirmed the observations made by existing works~\cite{proteus-eurosys17, tributary-atc18}: preemptions tend to be independent based on each individual spot market and each availability zone has a different and independent spot market\textemdash this is because each availability zone maintains capacity separately and therefore capacity preemptions in one zone are not associated with capacity preemptions in another. 

These observations motivate our design\textemdash even with 1-node redundancies, \system can recover from a majority of preemptions if consecutive nodes are not preempted at the same time; we maximize this possibility with a best-effort approach that makes consecutive nodes in each pipeline come from different zones. Although this may increase communication costs, it does not lead to visible performance impacts for \system because \system only sends (small amounts of) activations data between nodes.  



\MyPara{Strawman \#1: Checkpointing.} We next show why a technique based on checkpointing and restarting does not work.  
We developed a new checkpointing system on top of DeepSpeed~\cite{deepspeed-sc20},  providing checkpointing and restarting functionalities similar to TorchElastic~\cite{pytorch-elastic}. We modified DeepSpeed to checkpoint \emph{continuously} and \emph{asynchronously}. In particular, each worker moves a copy of any relevant model state to CPU memory whenever the state is generated; the CPU then asynchronously writes it to remote storage so that training and checkpointing can fully overlap. 
During restarting, our system automatically adapts the prior checkpoints to the new pipeline configurations.

To understand how well this technique performs, we used it to train GPT-2 over 64 p3.2xlarge GPU spot instances on EC2. We profiled the training process and collected the checkpointing times, reconfiguration overheads, and total execution time. Figure~\ref{fig:checkpoint_perf} reports these results. 
The blue sections represent the times the system spent making actual progress for training. The red sections represent the times on reconfiguring (\ie, restarting) while the orange sections show the times for wasted work\textemdash the computation that was done but not saved in checkpoints; the system ended up redoing these computations after restarting. This is because preemptions often occur during checkpointing, and hence, the system must roll back to a previous checkpoint. Frequent rollbacks slows down the training significantly. 
As shown, although checkpointing itself can be done efficiently, the restarting overheads (\ie, for adapting existing checkpoints to new pipeline configurations) and the wasted computations take \textbf{77\%} of the training time.

\begin{figure}[t]
  \centering
  \includegraphics[scale=0.6]{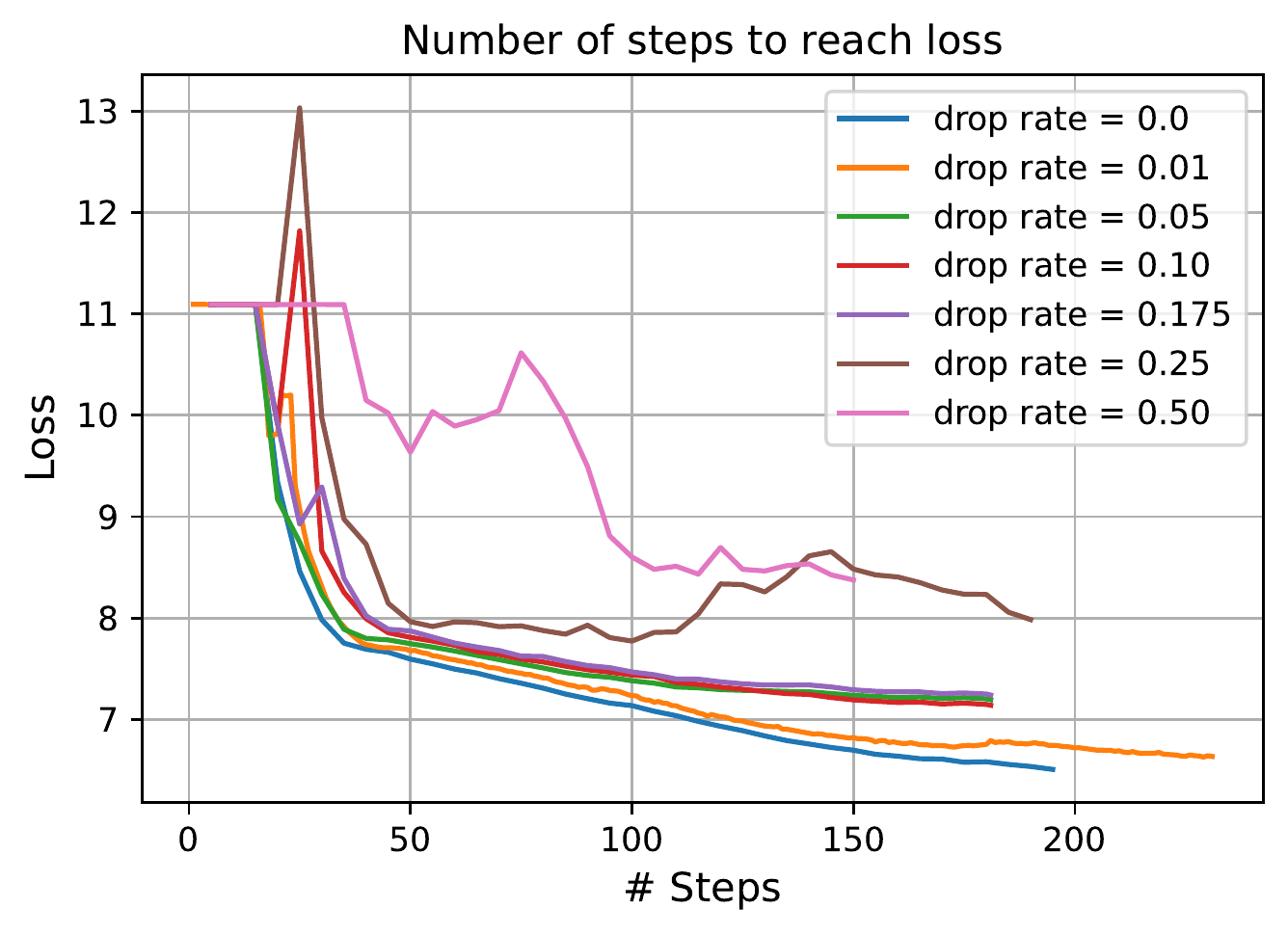}
  \caption{
    Effects of sample dropping under different rates.}
  \label{fig:sample_dropping}  
  \vspace{-1em}
\end{figure}

\MyPara{Strawman \#2: Sample Dropping.}
An alternative approach that has shown promise is to take advantage of the
statistical robustness of DNN training and allow some samples to be dropped so that
training can continue without significant loss of
accuracy~\cite{ieee/2018/data_dropout, corr/abs-1904-12043}. These techniques
are also known as \emph{elastic batching} because dropping
samples is equivalent to changing the effective batch size at a training
iteration (with the learning rate dynamically adjusted). 

In the case of pipeline parallelism, we implemented sample dropping by suspending a pipeline upon losing an instance while letting other data-parallel pipelines continue to run. The system performs optimizer steps with the gradients of
whichever data-parallel pipelines are able to complete that training step.
Learning rate was adapted linearly with respect to the effective batch size to make sure that the only effect on the accuracy
is the lost samples, but \emph{not} a mismatch between hyperparameters and training
configurations. In doing so, the training can continue for sometime without a
reconfiguration (which is needed upon allocations).

We conducted a set of experiments to simulate
the effect of sample dropping on model accuracy with a range of drop rates. Note that we could not obtain these results with the actual spot instances because we could not control the preemption rate. 
We ran a pre-training benchmark with GPT-2 using 16 on-demand instances from the same EC2 family, which form four data-parallel pipelines, each with four stages. 
To consider a range of different failure models, we
used different rates of preemption to generate preemption events. Upon a preemption event, we randomly selected a pipeline and zero out the pipeline's gradients in that iteration. 
We measured the model's evaluation accuracy every 5 training steps.
These results are shown in \figref{sample_dropping} where each curve represents the function of the number of steps needed to reach a given loss for a particular drop rate. 

Similarly to checkpointing, sample dropping works well for low preemption
rates, but when frequent preemptions occur, many
samples can be lost quickly and its impact on model accuracy quickly grows to be  too significant to overlook.
While this experiment was not an exact recreation of a sample
dropping scenario, these results represent an \emph{under-approximation} of the
effect of the actual sample dropping (which can lose more accuracy than reported by Figure~\ref{fig:sample_dropping}).
This is because 
the actual sample dropping rate should be higher than the instance preemption rate\textemdash a preempted instance would likely be down
for some time and consecutive samples would be dropped in a real setting.
Note that training samples are shuffled before
loading; hence, the effects of randomly dropping consecutive samples (\ie, the actual
scenario) and dropping random samples sporadically (\ie, our experiment) should be
similar.


\MyPara{Strawman \#3: Live Migration in Grace Periods.} Another potential approach is to use the grace period before each preemption to migrate data from the preempted node to a live node. However, this approach suffers from significant drawbacks. First, grace periods vary from cloud to cloud. While AWS spot instances have 2 minutes before preemption, GCP and Azure provide only 30 seconds, with GCP not even guaranteeing such a warning. For large models, this can be too short of a warning and may not leave sufficient time to save model updates into a checkpoint. 

A more important issue is that this approach depends on always maintaining enough idle nodes as migration targets. There is no way to guarantee that, unfortunately, with the current spot market. Even if we could over-provision and reserve a certain number of standby nodes, these nodes can also be preempted and it is impossible to ensure when a set of nodes in the pipeline are preempted, there are enough standby nodes for data migration. In fact, during our experiments, for each preemption event, the number of new allocations we could obtain was always less than the number of preempted nodes (as shown in Figure~\ref{fig:preemptions_64}).


\section{Overview}
\label{sec:overview}
\MyPara{Goal and Non-Goal.} Our goal is \emph{not} to automatically determine the cheapest way to train a given model (\eg, which parallelism model can lead to the largest cost savings). Instead, \system{} aims 
to enable efficient and preemption-safe training over cheap spot instances. 

\MyPara{User Interface.} To use \system, a user specifies two system parameters $D$ and $P$, as they normal would to use other pipeline-parallel systems, where $D$ is the number of data-parallel pipelines and $P$ is the pipeline depth. Due to the need of storing redundant layers, \system requires a larger pipeline depth $P$ than a normal pipeline-parallel system such as PipeDream~\cite{pipedream-sosp19}. We observed, empirically, that to avoid swapping data between CPU and GPU memory on the critical path, \system's pipeline should be $\thicksim$1.5$\times$ (see \S\ref{sec:overhead}) longer than an on-demand pipeline due to the extra memory needed to   
(1) hold the redundant layers and (2) accommodate potential pipeline adjustments. Given that spot instances are much cheaper (\eg, 3-4$\times$ on EC2) than on-demand instances, training with 1.5$\times$ more nodes still leads to significantly reduced costs. While we recommend 1.5$\times$ more nodes, the number of active instances in a cluster is often much smaller due to preemptions and incremental allocations. 

$P \times D$ will be the size of the spot cluster \system attempts to maintain throughout training.  Preemptions can cause \system to reduce the pipeline depth and/or the number of pipelines; in such cases, \system would request more instances to bring the size of the cluster back to $P \times D$. However, \system would never try to scale the training beyond $P \times D$. In other words, $P$ and $D$ are the \emph{upper bound} of the pipeline depth and number of pipelines.  It is important to note that the goal of the autoscaling framework we build for \system is to adjust the pipelines \emph{passively} in response to node preemptions and additions that we cannot control, rather than \emph{proactively} finding an optimal cluster configuration to achieve better performance. This distinguishes \system from existing works on autoscaling distributed training~\cite{elastic-mlsys20,varuna-2021,elastic-nsdi21}, whose goal was to find better configurations.



\begin{figure}[t]
  \centering
  \includegraphics[scale=.3]{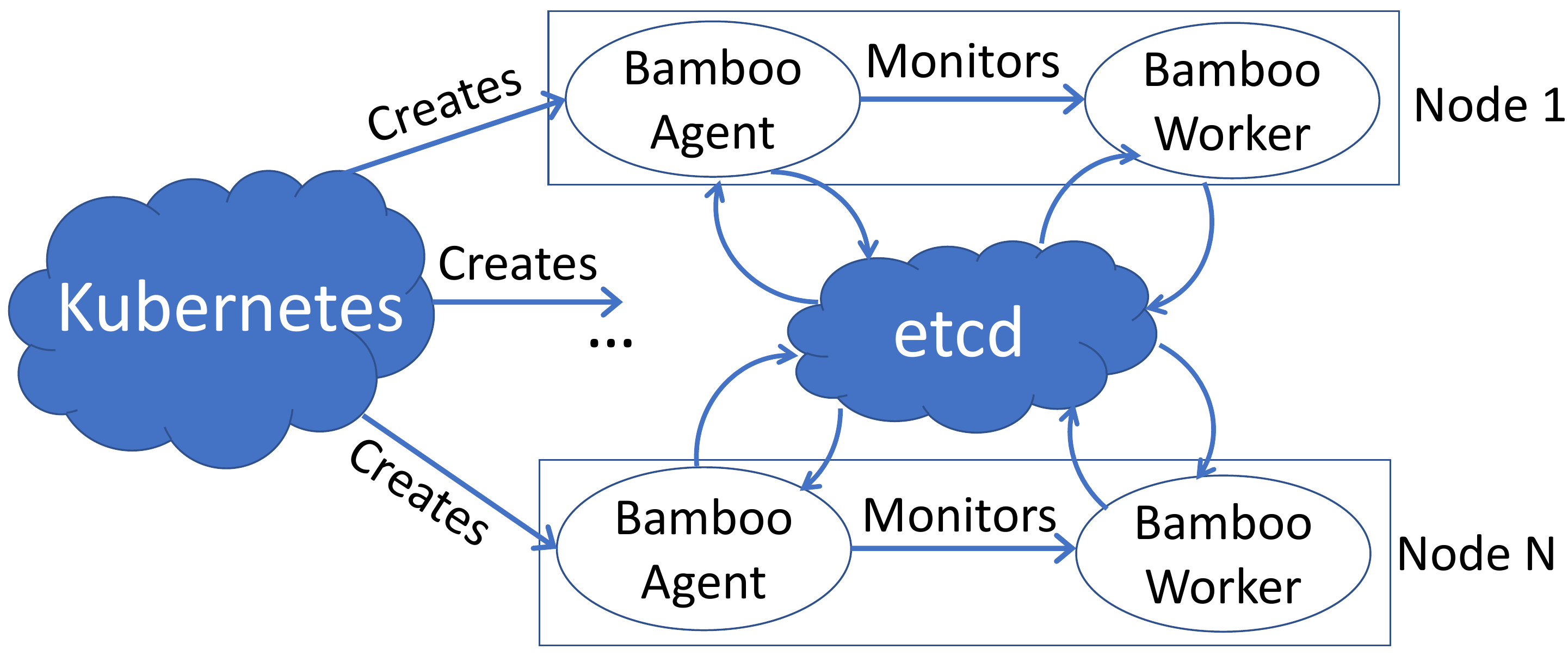}
  \vspace{-1em}
  \caption{
    \system{} runs one agent process per node (\ie, spot instance). 
    An agent monitors worker processes (each running a training script) that use our modified DeepSpeed. All workers and agents coordinate through \codeIn{etcd}~\cite{etcd}.
  }
  \label{fig:system-overview}
  \vspace{-.5em}
\end{figure}

\MyPara{System Overview.}
Figure~\ref{fig:system-overview} shows an overview of our system.
We built \system{} on TorchElastic~\cite{pytorch-elastic} and DeepSpeed~\cite{deepspeed-sc20}. In particular, we built the \system{} agent, which runs on each node to kill/add a data-parallel pipeline, on top of TorchElastic. The agent monitors a \system{} worker process on the same node, which is a DeepSpeed application enhanced with our support for redundant computation. \system workers run $D$ data-parallel pipelines that use an \codeIn{all-reduce} phase to synchronize weights at the end of each iteration. Our spot instances are managed by Kubernetes~\cite{kubernetes}, which is configured to automatically scale by launching a \system{} agent on each new allocation.
Our agents communicate and store cluster state on \codeIn{etcd}~\cite{etcd}, a distributed key-value
store. 

\begin{figure}[t]
  \centering
  \includegraphics[scale=.3]{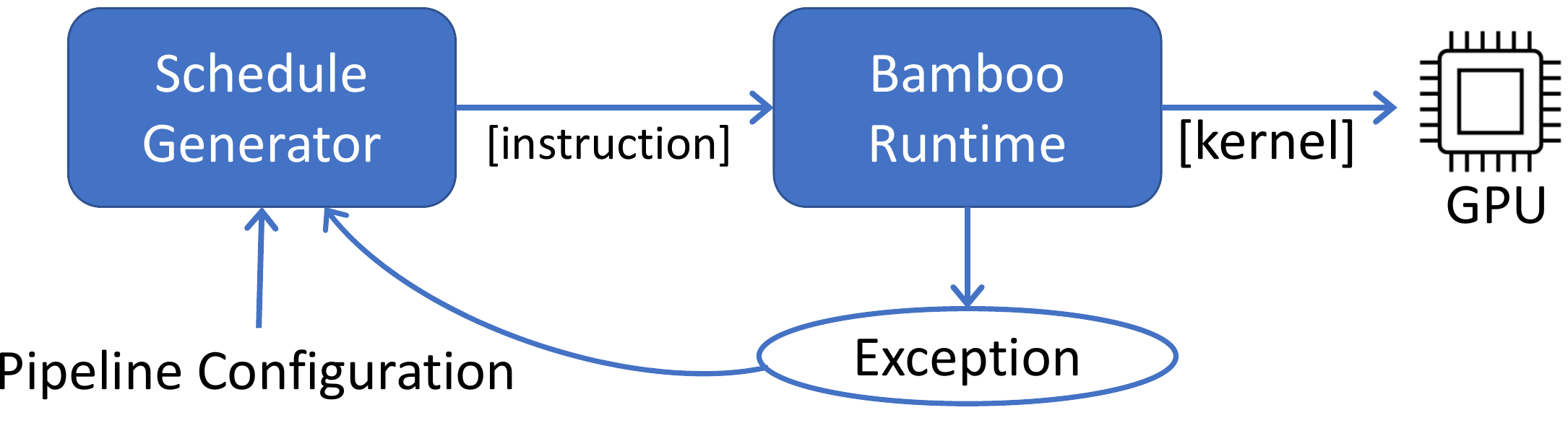}
  \caption{\system{} worker.}
  \label{fig:system-worker}
\end{figure}

Each \system{} worker uses a runtime to interpret the schedule, which produces a sequence of instructions, as shown in Figure \ref{fig:system-worker}. The schedule is generated statically based on the stage ID of the current worker and pipeline configurations, including the depth of pipeline and total number of microbatches. The instructions consist of a computation component (\ie, forward, backward, and apply gradient), and a communication component (\ie, send/receive activation, send/receive gradient, and all-reduce). 
The \system runtime interprets these instructions by launching their corresponding kernels on GPU. Communication instructions can fail due to preemptions. Upon a failure, the runtime throws an exception and falls back to use a failover schedule.

\section{Redundant Computation}
\label{sec:redundant-computation-and-recovery}
For ease of presentation, our discussion focuses on one node running one stage in the pipeline. Support for multi-GPU nodes will be discussed shortly.


Preemption of a node is detected by its neighboring nodes in the same pipeline during the execution of communication instructions. 
If a node on one side of the communication is preempted, the node on another side will catch an IO exception due to broken socket and update cluster state on \codeIn{etcd}. \system detects preemptions based on socket timeout. Although we could let a node to be preempted actively notify its neighbors in the grace period before the preemption, the length of this period varies across different clouds and hence \system does not use it currently.

Since the victim node communicates with two nodes in the pipeline, both of its neighbors can catch the exception. The observed exception will be shared between these two nodes through \codeIn{etcd}. This two-side detection is necessary for \system to understand which node fails and generate the failover schedule.
In addition to the two neighbors, nodes in other pipelines involved in the \codeIn{all-reduce} operation also need to be informed. To safely perform \codeIn{all-reduce}, each node participating in \codeIn{all-reduce} reads the up-to-date cluster state on \codeIn{etcd} and, if another pipeline has a failure, waits until the failure is handled.

\subsection{Redundant Layers and Computation
\label{sec:weight-sync}}

To quickly recover from preemptions, \system{} replicates the model partition on each worker node in each data-parallel pipeline. Instead of saving these replicas to a centralized remote storage (like checkpointing), \system takes a \emph{decentralized} approach by letting each node replicate its own model partition (\ie, layer shard) on its predecessor node in the same pipeline. The first node has its layer replica stored on the last node in the pipeline. Conceptually, the last node is considered the ``predecessor'' of the first node. For simplicity of presentation, we use \emph{forward stage IDs} to identify nodes, that is, a node that runs the forward stage $n+1$ is always considered as a successor of a node running the forward stage $n$ (although in the backward pass, $n+1$ is a stage before $n$).  

Our key idea is to let each node run normal (forward and backward) computation over its own layers and redundant (forward and backward) computation over the replica layers for its successor node. Let FRC$^{m}_{n}$/BRC$^{m}_{n}$ denote the forward/backward redundant computation that is performed on node $m$ for node $n$, respectively. In \system, $n = (m + 1) ~\mathit{mod}~ \mathit{P}$ where $P$ is the pipeline depth. 
Let FNC$_n$/BNC$_n$ denote the forward/backward normal computation on node $n$. In \system's pipeline, FRC$^{n}_{n+1}$/BRC$^{n}_{n+1}$ is exactly the same computation as FNC$_{n+1}$/BNC$_{n+1}$, working with the same model parameters and optimizer states. 
To enable the last node to perform RC for the first node, we let it fetch input samples directly. 

\begin{figure}[h!]
    \centering
    \includegraphics[scale=.6]{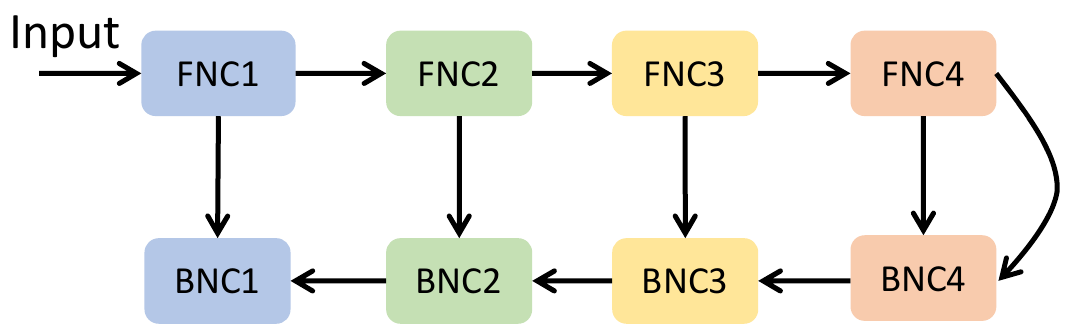}
    \caption{Dependencies between normal pipeline stages.
    \label{fig:dependencies}}
    \vspace{-1em}
\end{figure}

\MyPara{Why Neighboring Nodes?}
Due to our focus on pipeline parallelism, \system performs RC on predecessor nodes to exploit \emph{locality} for increased efficiency. To see this, we first need to understand the dependencies between different (backward and forward) pipeline stages that a microbatch goes through, as illustrated in Figure~\ref{fig:dependencies}. For each forward stage FNC$_n$, it depends only on the output of its previous stage FNC$_{n-1}$. However, for each backward stage BNC$_n$, it has two dependencies: one on the output of stage BNC$_{n+1}$ and a second on its corresponding forward stage FNC$_n$. The first is a \emph{hard} dependency without which BNC$_n$ cannot be done, while the second is a \emph{soft} dependency primarily for efficiency\textemdash intermediate results produced by FNC$_n$ can be reused to accelerate BNC$_n$. Without such cached results, BNC$_n$ has to recompute many tensors (\ie, tensor rematerialization~\cite{recompute-tianqi-16}), leading to inefficiencies.

\begin{figure}[t]
    \centering
    \includegraphics[scale=.51]{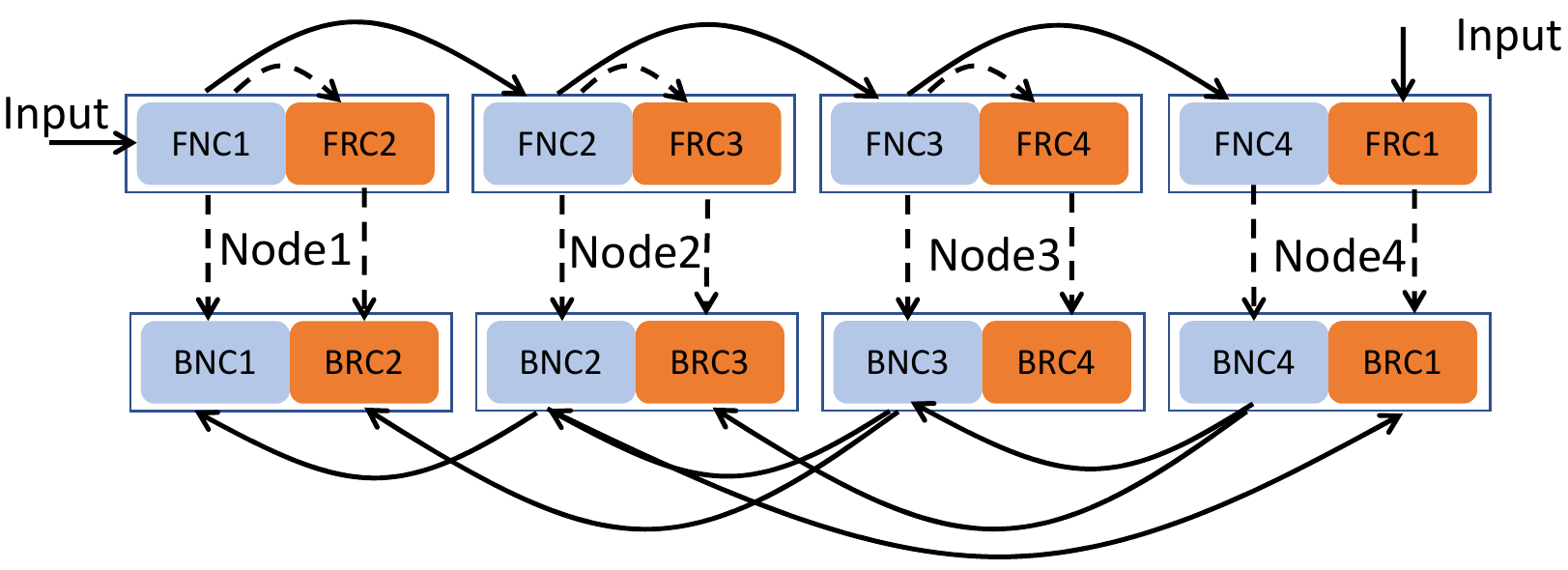}
    \vspace{-1em}
    \caption{Dependencies between RC-enabled pipeline stages: solid/dashed arrows represent inter/intra-node dependencies; for simplicity, FRC$_n$/BRC$_n$ in the figure represents FRC$^{n-1}_n$/BRC$^{n-1}_n$.
    \label{fig:rc-dependencies}}
\end{figure}

Figure~\ref{fig:rc-dependencies} shows dependencies on an RC-enable pipeline where each node performs both normal and redundant (backward and forward) computation. Here solid/dashed arrows represent inter/intra-node dependencies. By running FRC for node ${n+1}$ on node $n$, \emph{locality benefit} can be clearly seen because FRC only creates intra-node dependencies, which do not incur any extra communication overhead. However, in a backward pass, such a locality benefit does not exist for BRC$^{n}_{n+1}$, which requires the output of BNC$_{n+2}$ and incurs much extra communication. This motivates our eager-FRC-lazy-BRC design which does not perform BRC until a preemption occurs and hence eliminates the extra communication cost in normal executions. 

Note that we could also perform FRC lazily, but this would significantly increase the pause time for recovery. This is because (1) recovering from preemptions at both forward and backward pass now require a pause; and (2) lazy FRC would not produce intermediate results that can be used to speed up BRC and hence BRC's pause would be much longer. Since FRC can be scheduled in the pipeline bubble and overlap with FNC, performing it eagerly is a better choice. 


The careful reader may think of an alternative approach that places node $n$'s layer replica on node $n+1$ as opposed to node $n-1$ (\ie, its successor rather than its predecessor). This approach is symmetric to our design in that it turns inter-node dependencies for BRC into intra-node dependencies, but intra-node dependencies for FRC into inter-node dependencies. As a result, it eliminates the extra backward communication at the cost of increased forward communication. However, unlike \system's design that can use lazy BRC to eliminate the extra backward communication, it is not as easy to eliminate the extra forward communication with lazy FRC\textemdash if FRC is not done eagerly in each iteration, BRC (regardless of whether it is eager or lazy) must perform tensor re-materialization, which incurs a long delay. 

\MyPara{Level of Redundancy.} As with any redundancy-based systems, the more redundancies, the higher level of resilience. For example, since \system performs redundant computations only for one node, it cannot provide resilience when preemptions occur on consecutive nodes in a pipeline, in which case a reconfiguration is needed (see \S\ref{sec:reconfiguration-and-transfer}). However, enabling RC for multiple nodes can significantly increase the FRC time, making it much longer than what the bubbles can accommodate. Furthermore, the locality benefit (\ie, FRC only incurs intra-node dependency) does not hold anymore, because FRC now depends on the outputs of multiple nodes. This can slow down the training substantially. 

\MyPara{Takeaway.} Storing each node's replica layers on its predecessor and running eager-FRC-lazy-BRC achieves low-overhead RC for pipeline parallelism. While this design does not support consecutive preemptions, \system takes care to make consecutive nodes come from different zones. As discussed in \S\ref{sec:motivation}, if multiple preemptions occur at the same time, the preempted nodes are highly likely to be from the same zone. As a result, our node assignment reduces the chance of consecutive preemptions, making RC effective for most preemptions. Although cross-zone data transfer can incur an overhead, this overhead is negligible (\eg, $<$3\%), as reported in Appendix \S\ref{sec:cross}, because in pipeline-parallel training, each node only passes a small amount of activation data to its neighbors.

We refer to the preempted node as the \emph{victim node}, and the node saving the replica of the victim as its \emph{shadow node}.




\subsection{Schedule Redundant Computation}
\label{sec:schedule-rc}
It is straightforward to see that RC incurs an overhead in both time and memory.
We propose to (1) schedule FRC into the pipeline bubble to reduce forward computation overhead, (2) perform BRC lazily to reduce backward computation/communication overhead, and (3) offload unnecessary tensors to CPU memory to reduce memory overhead.

\begin{figure}[t]
    \centering
   
    \includegraphics[width=\linewidth]{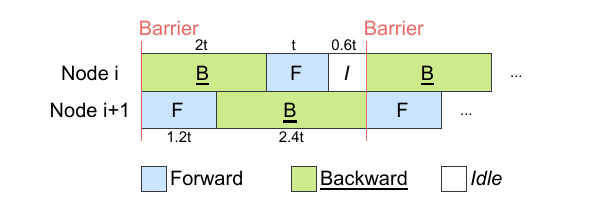}
     \vspace{-1em}
    \caption{A closer examination of the pipeline bubble. Here we assume the forward pass on node $i$ and $i+1$ takes time $t$ and $1.2t$, respectively. Hence, a bubble of 0.6$t$ exists before each communication barrier.
    \label{fig:instr-group}}
    \vspace{-1em}
\end{figure}

\MyPara{Eager FRC.} As discussed in \S\ref{sec:background}, the pipeline bubble can come from either imperfect scheduling or unbalanced pipeline partitioning. To illustrate, consider Figure~\ref{fig:instr-group} with PipeDream's 1F1B schedule. Suppose there are two consecutive nodes in the pipeline where both the forward and the backward computation of node $i+1$ run 1.2$\times$ slower than those of node $i$. The communication between these two nodes serves as a barrier. Since node $i$ runs faster, it always reaches the barrier earlier and waits there until node $i+1$ arrives. This wait period is where we should schedule FRC.

\system builds on the 1F1B schedule (Figure~\ref{fig:pipeline}(a)) due to its additional efficiency compared to GPipe's schedule (Figure~\ref{fig:pipeline}(b)).
However, even for 1F1B, bubbles widely exist in a pipeline\textemdash as a microbatch passes different pipeline stages, the later a stage, the longer the (backward and forward) computation takes. This is because for the 1F1B schedule, the number of active microbatches in a later stage is always smaller than that in an earlier stage. In Figure~\ref{fig:pipeline} (c), for example, node 1 has 3 active microbatches while node 2 only has 2. Consequently, later stages often consume less memory. To balance memory usage, the layer partition on a later node is often larger that that on an earlier node in the pipeline, and hence a later stage runs slower.

\begin{figure}[h!]
    \centering
    \vspace{-1.5em}
    \begin{adjustbox}{max width=\linewidth}
    \includegraphics{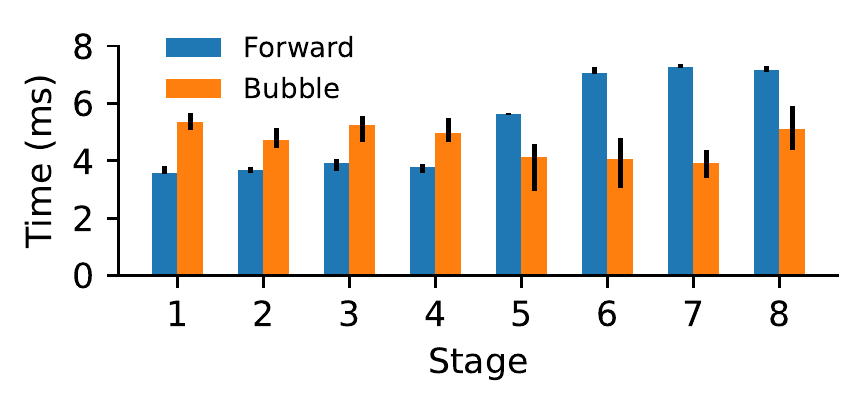}
    \end{adjustbox}
    \caption{Comparison between bubble size and forward computation. \label{fig:bubble}}
\end{figure}

\MyPara{Pipeline Size.} To understand whether pipeline bubbles are sufficient for running RC, we first measure the sizes of these bubbles and the forward computation for BERT, a typical language model, when it runs on 32 on-demand V100 GPU instances (\ie, four data-parallel pipelines each with eight pipeline stages). We manually insert a barrier before each peer-to-peer communication, treating the time spent on its corresponding NCCL kernel as a bubble. These results are reported in Figure \ref{fig:bubble}.

To make memory evenly distributed across stages, more layers are placed on the last few stages and hence the amount of forward computation grows with the stage. In this pipeline, for the first 4 stages, the bubble time is long enough to fit the entire FRC (\ie, the bubble at stage 1 should run the forward computation for stage 2). For the last 4 stages, the bubble time is shorter than the forward computation time\textemdash it can still cover $\thicksim$60\% of its FRC. The rest of the FRC on these nodes is run in parallel with their regular forward computation, as discussed shortly in this section.

\MyPara{Scheduling.} Based on this observation, we schedule FRC on a node before the node starts communicating with its successor node. This is where a bubble exists.
In cases where the FRC cannot fit entirely into the bubble (\ie, for the last four stages in Figure~\ref{fig:bubble}), we overlap FRC and FNC as much as we can. However, for the same microbatch, FRC$^n_{n+1}$ depends on FNC$_n$ and they cannot run in parallel. To resolve this dependency issue, we focus on different microbatches for FNC and FRC. That is, \system schedules FNC$_n$ for the $k$-th microbatch and FRC$^n_{n+1}$ for its previous ($k-1$)-th microbatch to run in parallel. Since there is no dependency between them, their executions can overlap.


To reduce memory overhead, \system follows a well-known principle to offload less frequently used tensors to CPU memory. Specially, since BRC is \emph{not} performed in normal training passes, FRC's outputs and intermediate results are not needed until a preemption occurs and BRC is triggered. As a result, we swap out these data after FRC is done for each microbatch on each node. These data take the majority of FRC's memory consumption; swapping them out significantly reduces FRC's GPU memory usage~\cite{rajbhandari2020zero}. However, we leave the redundant weights in GPU memory for efficient FRC because these weights are needed for FRC on each microbatch.  

 \begin{figure}[t]
     \centering
     \includegraphics[width=\linewidth]{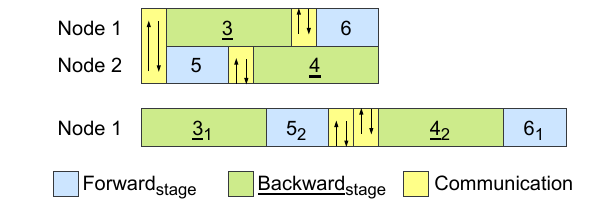}
     \vspace{-1.5em}
     \caption{A example of merged instruction sequences in failover schedule. We use PipeDream's 1F1B schedule as shown Figure~\ref{fig:pipeline}(c), and assume node 2 is the victim node and node 1 is the shadow node.}
     \label{fig:instr-merge}
     \vspace{-.5em}
 \end{figure}
 
\MyPara{Lazy BRC and Recovery.}
BRC is executed by a \emph{failover schedule} which a node runs when detecting its successor node fails. In particular, for the current iteration, all the lost gradients must be re-computed, while for the following iterations, all instructions of the victim node must be executed by its shadow node (until a reconfiguration occurs). Nodes that originally communicate with the victim node are transparently rerouted to the shadow node.  The failover schedule is generated by merging the schedules of the victim and shadow node.
In particular, a schedule consists of a sequence of instructions and we divide it into two groups\textemdash  (1) continuous communication instructions, which is placed at the head of a group and (2) computation instructions that can be executed without remote data dependencies.

When the two instruction groups (from the victim and shadow nodes) are merged, the instructions are interleaved with the following rules.
(1) Communication instructions are still placed in the beginning of the merged groups.
(2) Communications that used to be inter-node between the victim and the shadow are removed.
(3) External communications from the victim node are first performed.
(4) Computation instructions are ordered such that backward computation is always executed earlier; after the backward computation is done, the memory occupied by intermediate results is freed.
Figure \ref{fig:instr-merge} shows an example of merged instruction sequences if node 2 is the victim node and node 1 is the shadow node.

\MyPara{Support for Multi-GPU Nodes.}  \system{}’s RC works for multi-GPU settings\textemdash this requires replicating all layers that belong to the GPUs of one node in the GPUs of its predecessor node. In other words, we use ``group replicas'' as opposed to individual replicas. However, in the presence of frequent preemptions, using multi-GPU would yield poorer performance\textemdash losing one node (with multiple GPUs) is equivalent to losing multiple nodes in the single-GPU setting. Our evaluation (\S\ref{sec:evaluation}) shows that it is much harder to allocate new multi-GPU nodes during training than single-GPU nodes. 

Once \system{} loses too many nodes or there are many idle nodes (\ie, new allocations) waiting to join the pipelines, \system{} launches a reconfiguration. Details of the reconfiguration process can be found in \S\ref{sec:reconfiguration-and-transfer}.

\MyPara{Support for Pure Data Parallelism \label{sec:data}}
\system supports pure data parallelism (without model partitioning). Due to space constraints, here we briefly discuss how it is supported. We use the same redundant computation strategy\textemdash Bamboo replicates the parameter and optimizer state of each node on a different node and uses these replicas as redundancies to provide quick recovery. For pure data parallelism, there is no bubble time to schedule RC. Eager FRC would be equivalent to overbatching (\ie, each node processes its original minibatch plus a redundant minibatch). To reduce the FRC overhead and make RC fit into the GPU memory constraints, we over-provision spot instances (by 1.5$\times$, in the same way as discussed in \S\ref{sec:redundant-computation-and-recovery}) to make each node process a smaller batch. 

Enabling eager FRC doubles the batch size. However, it results only in a $\thicksim$1.5$\times$ increase in the computation time due to the parallelism provided by GPUs. This overhead can be effectively reduced by slightly over-provisioning (1.5 $\times D$) nodes, increasing the degree of parallelism and decreasing the impact of overbatching. This enables us to run FRC eagerly without incurring much overhead (\ie, <10\%).

\section{Pipeline Reconfiguration
\label{sec:reconfiguration-and-transfer}}

Reconfiguration introduces a much longer pause to the training process than recovering using RC.
The goal of reconfiguration is to rebalance pipelines so they can withstand
more failures as training progresses and continue to yield good performance. 
Reconfiguration also attempts to allocate more instances to maintain the cluster size. As shown in \S\ref{sec:motivation}, asynchronous checkpointing is very efficient (but frequent restarting is not), and hence, \system periodically checkpoints the model state. These checkpoints will not be used unless \system restarts the training from a rare fatal failure (\ie, too many nodes are preempted so that training cannot continue).


\MyPara{Reconfiguration Triggering.}
Reconfiguration is triggered immediately when  (1) consecutive preemptions occur simultaneously and (2) \system determines that there is an urgent need to rebalance the pipelines at the end of an optimizer step. To do (2), the workers retrieve the cluster state from \codeIn{etcd}, allowing them to see
how many preemptions have occurred and in which pipeline they have occurred.
They can also see how many workers are currently waiting to join the next rendezvous.

There are two main conditions for triggering reconfiguration at the end of an optimizer step:
(a) the cluster has gained enough new nodes to reconstruct a new pipeline, and (b)
\system has encountered many preemptions and is close to a critical failure in the next step (\eg, encountering another preemption would cause us to suspend training), in which case we must pause the training to allocate more nodes.


\MyPara{Reconfiguration Policy.}
\system attempts to maintain the pipeline depth $P$ specified by the user.
Therefore, our top priority at a reconfiguration is to reestablish a full pipeline of
depth $P$. In this case, if we have had $F$ failures and $J$ ($>F$) nodes are waiting to join the cluster (\ie, new allocations arrive as \system runs on the ``spare tire''), we can fully recover all pipelines to depth $P$. The remaining ($J-F$) nodes are placed in a standby queue to provide quick replacement upon future failures. 
However, if the number of nodes joining is smaller than $F$, we may end up having a number of $N$ nodes such that $N \% P \neq 0$. In this case, instead of creating asymmetric pipelines (which complicates many operations), we move some nodes into the standby queue and decrease the total number of data-parallel pipelines. 
A final case is that the number of nodes joining, together with those in the standby queue, can form a new pipeline, and in this case we add a new pipeline to the system. In all these cases, the redundant layers are redistributed among the set of nodes participating in the updated pipelines.

\MyPara{How to Reconfigure.}
Once a reconfiguration is triggered, each node must be assigned a new stage (with new layers, state, and redundancies); it also needs to figure out if it will need to send or receive model
and optimizer state from other nodes.
Whichever nodes hits the rendezvous barrier first decides the new cluster
configuration and puts the decision on \codeIn{etcd} for all other nodes to read.
To minimize the amount of data sent in layer transfer, \system transfers layers in such a way that each node can reuse its old model and optimizer state as much as possible.







\section{Evaluation}
\label{sec:evaluation}

\begin{table}[t]
\small
  \centering
  \begin{adjustbox}{max width=\linewidth}
  \begin{tabular}{lllll}
      \rowcolor[HTML]{EFEFEF} 
    Model & Dataset & $D$ & $P$ \\
  ResNet-152~\cite{resnet} & ImageNet~\cite{imagenet-cnn-cacm17}& 4 & 8$\times$1.5 (12) \\ 
 VGG-19~\cite{vgg} & ImageNet~\cite{imagenet-cnn-cacm17} & 4 & 4$\times$1.5 (6) \\ 
 AlexNet~\cite{imagenet-cnn-cacm17} & Synthetic data & 4 & 4$\times$1.5 (6) \\\hline
 GNMT-16~\cite{gnmt} & WMT16 EN-De & 4 & 4$\times$1.5 (6) \\
     BERT-Large~\cite{bert} & Wikicorpus En~\cite{bert} & 4 & 8$\times$1.5 (12) \\ 
     GPT-2~\cite{gpt2} & Wikicorpus En~\cite{bert} & 4 & 8$\times$1.5 (12) \\ 
 \end{tabular}
 \end{adjustbox}
   
  \caption{Our models, datasets, pipeline configurations. \label{tab:benchmarks}}
  \vspace{-1em}
\end{table}

\system is implemented in $\thicksim$7K LoC as a standard Python library. 
We evaluated \system{} by pretraining a range of popular vision and language models, as shown in Table~\ref{tab:benchmarks}. We used two tasks and four datasets in our
experiments: (1) image classification, using the ImageNet-1K
(ILSVRC12)~\cite{imagenet-challenge} dataset and (2) translation, using the WMT16 English to German dataset for GNMT-16 and the Wikicorpus dataset~\cite{bert} for BERT and GPT-2. 
For the first four (smaller) models that were also used in PipeDream~\cite{pipedream-sosp19} (which actually used smaller versions of these models), 
we took the values of $D$ (the number of data-parallel pipelines) and $P_{demand}$ (pipeline depth) from PipeDream~\cite{pipedream-sosp19}'s configurations.

As discussed earlier in \S\ref{sec:overview}, to avoid swapping \system needs 1.5$\times$ more instances for each pipeline and hence each $P$ reported in Table~\ref{tab:benchmarks} equals 1.5$\times P_{demand}$. For BERT and GPT2, we used 4 and 8$\times$1.5=12 as $D$ and $P$. We have also evaluated with another pipeline depth $P_h=P_{demand}\times\frac{Price_{demand}}{Price_{spot}}$; these results can be found in \S\ref{sec:scaling}.

We trained these models on a spot cluster from EC2's p3 family where each instance has V100 GPU(s) with 16GB GPU memory and 61GB CPU memory. Each on-demand instance costs \$3.06/hr per GPU while the price of its spot counter-part (at the time of our experiments) is \$0.918/hr. Our evaluation uses two on-demand baselines: (1) p3 instances each with four V100 GPUs (Demand-M) and (2) p3 instances each with a single GPU (Demand-S). For both baselines, the pipeline configuration was the same and all nodes were obtained from one availability zone.


For all experiments, we trained each model to a target validation accuracy, which is a particular number of samples for the model. We did not train them to higher accuracies because large models take a huge amount of time to train (\eg, weeks) to reasonable accuracies; using such a large amount of resources (even spot instances) goes beyond our financial capabilities. Furthermore, \system uses synchronous training where the time per iteration is fixed; hence, training for extended time would not change our results.

For on-demand instances, we used the largest per-GPU minibatch that fits in one GPU’s memory\textemdash anything larger
yields out-of-memory exceptions. This ensures that we hit peak
achievable FLOPs on a single device. For data-parallel runs with $n$
workers, the global minibatch size is $n\times g$ where $g$ is the minibatch size. The global minibatch sizes we used are consistent with those used by the ML community
and reported in the literature for these models. We used a per-GPU
minibatch size of 256 per GPU for VGG-19, 512 for AlexNet, 2048
for ResNet-152, 32 for GNMT-16, 256 for BERT-Large, and 256 for GPT-2. For microbatch size, we always selected a small value and tuned it for different models/configurations. 
We trained the vision models with an initial
learning rate of 0.001, respectively, with a vanilla SGD optimizer~\cite{kiefer-sgd-1952am}. For language models, we used the Adam optimizer~\cite{kingma2014adam} with an initial learning rate of $6e^{-3}$. We used half (fp16) precision in all our experiments.

\begin{table}[t]
\begin{adjustbox}{max width=\linewidth}
\small
  \centering

  \begin{tabular}{ll|rrr}
  \rowcolor[HTML]{EFEFEF}  Model & System & Throughput & Cost (\$/hr) & Value \\\hline
  \multirow{2}{*}{ResNet}  & Demand-M  &  30  & 97.92      & 0.31 \\
                           & Demand-S  &  32  & 97.92   &  0.33  \\ 
                           & \system-M  &  [19.35, 15.69, 8.22]   &   [44.33, 40.01, 37.21]       & [0.43, 0.39, 0.22] \\
                           & \system-S  &  [\textbf{21.67}, 19.41, 12.13] & [\textbf{42.23}, 40.39, 36.72]  &  [\textbf{0.51}, 0.48, 0.33] \\\hline
                    
 \multirow{2}{*}{VGG}          & Demand-M  & 197 &  48.96    & 4.02 \\
                               & Demand-S  & 167 & 48.96   & 3.41 \\
                               & \system-M  &  [93.34, 75.75, 64.22]      &     [21.31, 19.55, 18.43]        & [4.38, 4.11, 3.48] \\
                               & \system-S  & [\textbf{153.31}, 124.88, 98.21]  & [\textbf{20.19}, 19.28, 18.36] & [\textbf{7.59}, 6.48, 5.35] \\\hline
                               
 \multirow{2}{*}{AlexNet}      & Demand-M  & 359  & 48.96 & 7.33 \\ 
                               & Demand-S  & 336  & 48.96 & 6.86 \\
                               & \system-M  &   [271.06, 207.43, 143.57]     &    [21.31, 19.55, 18.43]         & [12.72, 10.61, 7.79]  \\
                               & \system-S  & [\textbf{340.32}, 321.65, 280.42] & [\textbf{20.19}, 19.28, 18.36] & [\textbf{16.86}, 16.68, 15.27] \\\hline
                               
 \multirow{2}{*}{GNMT}         & Demand-M  &  27      &     48.96     & 0.55 \\ 
                               & Demand-S  & 24 & 48.96 & 0.49 \\
                               & \system-M  &   [13.95, 10.82, 6.33]    &    [21.31, 19.55, 18.43]         & [0.65, 0.55, 0.34] \\
                               & \system-S  & [\textbf{18.92}, 16.31, 8.8] & [\textbf{20.19}, 19.28, 18.36] &  [\textbf{0.94}, 0.85, 0.48] \\\hline
                               
 \multirow{2}{*}{BERT}         & Demand-M  &   118     &     97.92     &  1.21  \\ 
                               & Demand-S  & 108  & 97.92 & 1.10  \\
                               & \system-M  &  [71.22, 56.41, 41.68]     &      [44.33, 40.01, 37.21]       & [1.61, 1.41, 1.12] \\
                               & \system-S  & [\textbf{98.87}, 83.70, 60.59] & [\textbf{42.23}, 40.39, 36.72] & [\textbf{2.34}, 2.07, 1.65] \\\hline
 \multirow{2}{*}{GPT}          & Demand-M   &     32 &     97.92   &  0.32  \\ 
                               & Demand-S   & 30 & 97.92 & 0.30 \\
                               & \system-M  &   [17.73, 14.00, 11.54]    & [44.33, 40.01, 37.21]  &  [0.40, 0.35, 0.31]  \\
                               & \system-S & [\textbf{29.92}, 22.68, 13.78] & [\textbf{42.23}, 40.39, 36.72] & [\textbf{0.71}, 0.56, 0.38] \\[-1em]
 \end{tabular}
 \vspace{-.5em}
  \end{adjustbox}
  \caption{Comparisons between training with DeepSpeed over on-demand instances and \system over spot instances; throughput is defined as the number of samples per second. For Bamboo, we train each model three times, and their results are explicitly listed in the form of [$a,b,c$] for the 10\% (average), 16\%, and 33\% preemption rates, respectively. 
  \label{tab:overall}}
  \vspace{-.5em}
\end{table}

\subsection{Training Throughput and Costs \label{sec:overall}}


\MyPara{Overall Performance.}
To thoroughly and deterministically evaluate \system{}'s performance over spot instances under different preemption rates, we first ran a 48-node cluster (\ie, the configuration for ResNet, BERT, and GPT) and a 32-node cluster (\ie, for VGG, AlexNet, and GNMT) on AWS and collected a 24-hour preemption trace for each. On these traces, the \emph{hourly preemption rate} varies significantly, ranging from no preemption all the way to 16 nodes preempted (33\%), with an average rate of 4-6 nodes per hour (8-12\%). To account for such changes, we extracted from each trace three segments, each with a different hourly preemption rate: 10\%, 16\%, and 33\%. We used AWS' fleet manager to trigger preemptions by replaying these segments. Note that if we were to run \system over the uncontrolled spot cluster, there would be no way to enable a direct comparison. 

We trained ResNet, BERT, and GPT by replaying the three segments from the 48-node trace, and VGG, AlexNet, and GNMT by using the segements from the 32-node trace. 
These results are reported in Table~\ref{tab:overall}. 
In addition to the time and monetary costs, we used a metric called \emph{value}, which measures performance-per-dollar. Value is computed as $V$ = $\frac{T}{C}$ where $T$ is the training throughput, measured in terms of the number of samples per second, and $C$ is the monetary cost per hour. Throughout the evaluation, we used both value and throughput as our metrics.

Our first observation is Demand-M slightly outperforms Demand-S due to reduced cross-node communication. However, the difference is marginal as the amount of data (\ie, only activations) transferred over the network is small. 
\system{}-S significantly outperforms \system{}-M (\ie, \textbf{1.4$\times$} higher throughput and \textbf{1.5$\times$} higher value) because (1) multi-GPU nodes are subject to more GPU failures with the same number of preemptions and (2) it is much harder to to allocate new nodes in a timely fashion.

For \system{}-S, the results in each bracket of the form [$a$, $b$, $c$] show \system{}'s performance under the three preemption rates. The higher the preemption rate, the worse \system{}'s throughput and value. Given that the average preemption rate is $\thicksim$10\%, the first number in each bracket (highlighted) represents \system{}'s performance on the used spot cluster. On average, \system{}'s throughput (under the 10\% preemption rate) is  
15\% lower than DeepSpeed running over $D\times P_{demand}$ instances.  There are three major reasons. 
 
 First, the number of active instances in the spot cluster is actually lower than the requested size $D\times P$. For ResNet, for example, the average number of  instances throughout the training is only 25.58 although the requested cluster size is 48 (and the on-demand cluster always has 32 nodes).  The autoscaling group keeps attempting to add new instances but the total number of active instances only reaches the requested size for a small period of time.  
 
 Second, \system's reconfiguration contributes to reduced throughput\textemdash these overheads vary with environments and take an average of 7\% of the total training time. 
 
 Third, the time for each iteration increases due to eager FRC. This is the major source of overhead for language models such as GPT-2. A detailed evaluation of RC's overhead can be found in \S\ref{sec:overhead}. 
 
 Despite the small throughput reduction, \system delivers an overall of \textbf{1.95$\times$} higher value compared to training with on-demand instances. The benefit in value remains clear for five models (ResNet, VGG, AlexNet, BERT and GPUT) even when the preemption rate increases to 33\% (\ie, the worst-case segment of the collected trace).

\begin{figure*}[t]
  \centering
  \begin{tabular}{cccc}
  \includegraphics[scale=.55]{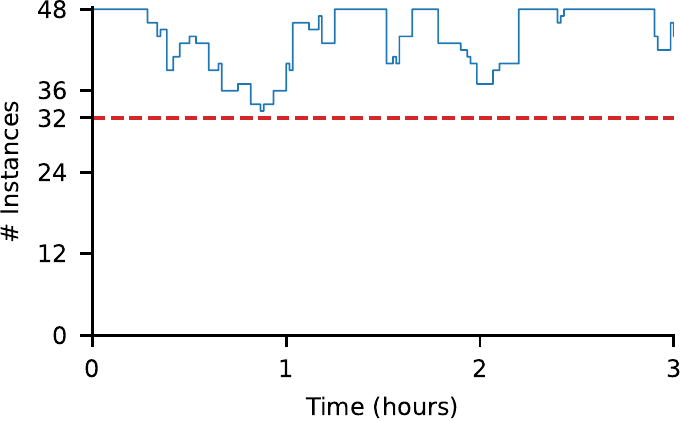} &
  \includegraphics[scale=.55]{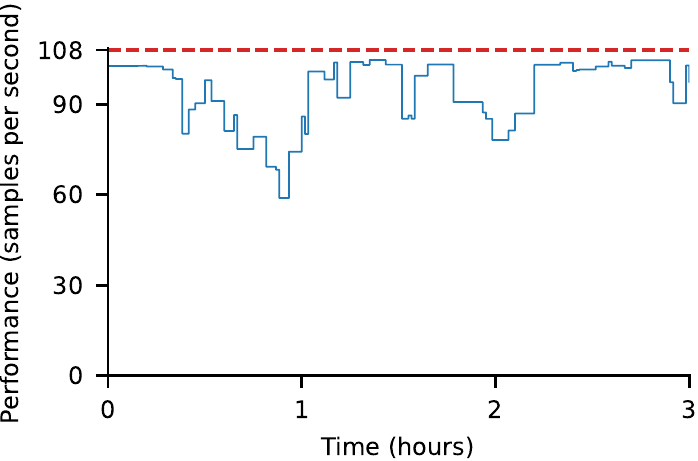} & 
   \includegraphics[scale=.55]{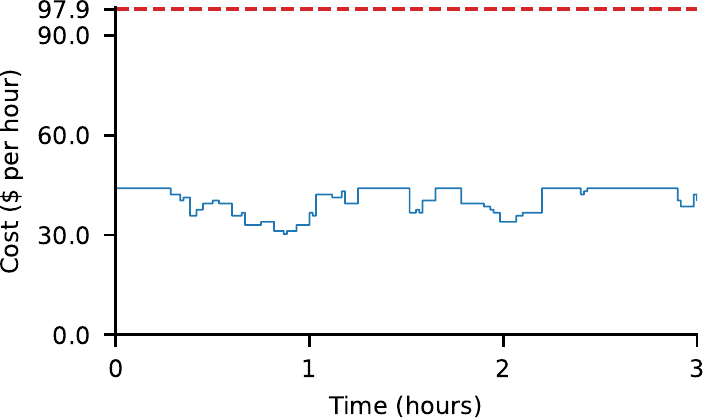} &
    \includegraphics[scale=.55]{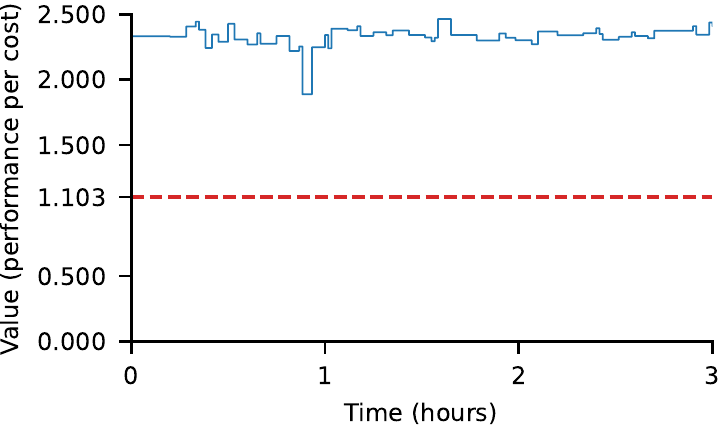} \\
    \includegraphics[scale=.55]{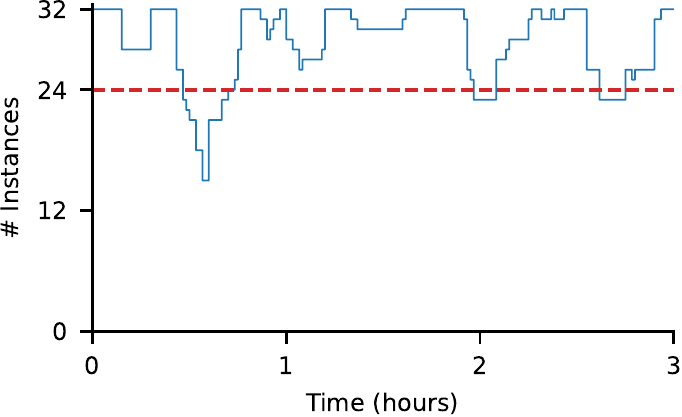} &
  \includegraphics[scale=.55]{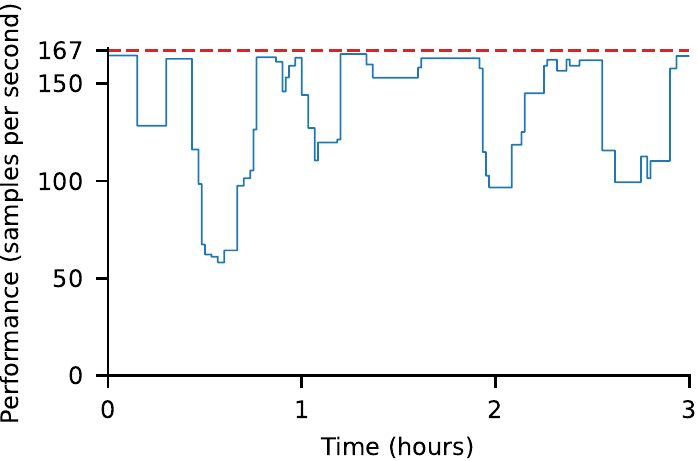} & 
   \includegraphics[scale=.55]{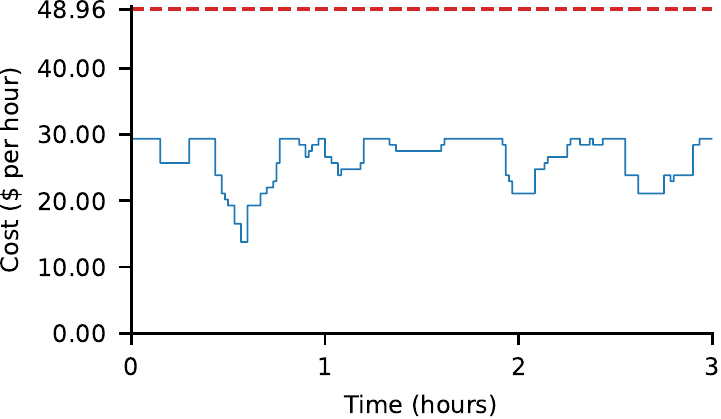} &
    \includegraphics[scale=.55]{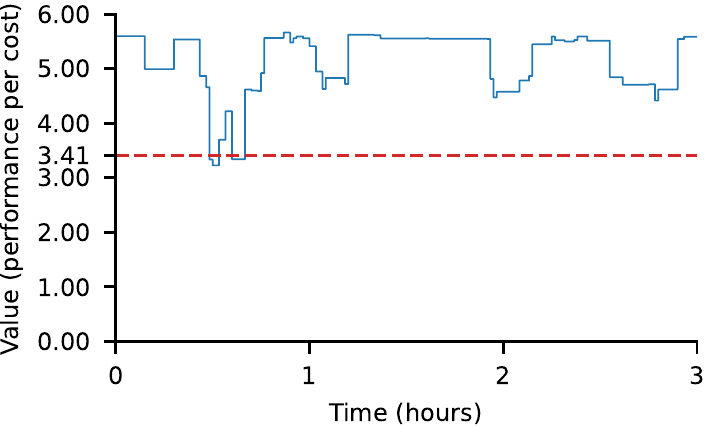} \\
   (a) Trace & (b) Training Throughput & (c) Monetary Cost & (d) Value
  \end{tabular}
  \caption{
    \system's training performance for BERT (top) and VGG (bottom), compared to on-demand instances (red lines).  \label{fig:training-detail}
  }
\end{figure*}

To have a closer examination of \system{}-S' training, we showed the traces for BERT-large and VGG-19, and plotted them in Figure~\ref{fig:training-detail}. The two rows show (a) preemption traces (under the 10\% rate), (b) training throughputs, (c) monetary costs, and (d) values, for BERT-large and VGG-19, respectively.  
Since \system{}-M underperforms \system{}-S, we focus on \system{}-S in the rest of the evaluation.


\begin{table*}[t]
  \centering
  \scriptsize

  	\begin{subtable}[l]{.65\textwidth}
  \begin{tabular}{rrrrrrrrr}
     \rowcolor[HTML]{EFEFEF} 
    Prob. & Prmt (\#) & Inter. (hr) & Life (hr) & Fatal Fail. (\#) & Nodes (\#) & Thruput &  Cost (\$/hr) &   Value \\
       0.01 &        8.50 &          2.08 &     15.20 &           0.06 &     45.18 &       87.99 &     41.11 &      2.10 \\
       0.05 &       48.15 &          0.44 &     10.14 &           0.23 &     43.65 &       76.35 &     39.73 &      1.90 \\
       0.10 &       99.77 &          0.23 &      6.71 &           0.29 &     41.69 &       72.12 &     37.94 &      1.88 \\
       0.25 &      276.52 &          0.10 &      3.13 &           1.04 &     35.80 &       60.12 &     32.58 &      1.82 \\
       0.50 &      709.83 &          0.06 &      1.49 &           5.98 &     26.96 &       40.37 &     24.53 &      1.59 \\
  \end{tabular}
   \caption{Results of simulating training BERT \emph{until completion}; each preemption
  probability ran 1,000 times.  \label{tab:simulation}}
    \end{subtable}
     	\hfill \hspace{-2em}
  	\begin{subtable}[r]{.25\textwidth}
  \begin{tabular}{rrrr}
 \rowcolor[HTML]{EFEFEF} 
    Prob. & Thruput &  Cost (\$/hr) & Value \\
       0.01 &   54.87 &     90.73 &      0.60 \\
       0.05 &   50.66 &     87.43 &      0.58 \\
       0.10 &   49.18 &     83.23 &      0.59 \\
       0.25 &   40.59 &     71.24 &      0.57 \\
       0.50 &   26.24 &     53.05 &      0.49 \\
       \end{tabular}
       \caption{Simulation results of training BERT-large with pipeline depth $P_h$ (which is 3.3$\times P_{demand}$).\label{tab:big-model}}
  \end{subtable}
  \vspace{-1.5em}
 \caption{Simulation results for more configurations.
 }
  \vspace{-.5em}
\end{table*}


\subsection{Different Failure Models\label{sec:scaling}}

The previous section demonstrated \system{} running on real spot instances. 
This section demonstrates \system{}'s ability to affordably train large models across a wide range of failure models. To this end, we developed an offline simulation framework that takes as input (1) the preemption probability (including preemption frequency and the number of preemptions in each bulk), (2) per-iteration training time, and (3) \system's recovery and reconfiguration time, automatically calculating training performance, costs, and values.  Here we focus on BERT-large and simulated its training until completion. 

We experimented using 5 different preemption probabilities (\ie, preemption
rate per hour), and kept the preemption probability constant throughout the
entire run (as opposed to replaying traces). To mimic realistic spot instance creation and preemption, we randomly
generated different creation probabilities per hour and also randomly picked zones for allocations. For each preemption probability, Table~\ref{tab:simulation} reports the average numbers of preemptions, intervals (\ie, average time, in hours, between preemption events), average lifetime of an instance (in hours), average numbers of fatal failures (which require a restart from a checkpoint), average numbers of instances in the cluster, throughput (\ie, 
\#samples per second), costs, and values, across 1,000 simulations.

Our simulations show that \system's values match our real-world runs as just reported in \S\ref{sec:overall}. Further, 
regardless of the preemption probability, the value of \system 
remains stable and is constantly higher than that of training with on-demand instances (which is 1.1). This is because most preemptions can be quickly recovered without introducing much overhead. The higher the preemption probability, the less the active instances running training jobs; this is the major source of the performance slowdown. However, the cost is reduced also proportionally, leading to stable values.  



\MyPara{Simulation for $P_h$.} To understand the tradeoff in choosing $P$, we experimented with another value of $P$ for BERT-large: $P_h$, which is $\frac{3.06}{0.918} \times P_{demand}$. This configuration 
represents the \emph{upper-bound} of the spot training resources that can be obtained within the cost of training with $P_{demand}$ on-demand instances (while $D$ remains unchanged). Note that in practice the number of active instances can barely reach the requested size and hence the cost of using a spot cluster of size $P_h\times D$ is often still much lower than training with an on-demand cluster of size $P_{demand}\times D$. 

To avoid incurring a large monetary cost, we used the same simulator to run this experiment. These results are reported in Table~\ref{tab:big-model}. As shown, using $P_h$ actually decreases both throughput (compared to 84 under $P$ in Table~\ref{tab:overall}) and value (due to significantly increased costs). This is because using too a large pipeline leads to poorer partitioning, underutilized resources and inferior performance. 

\subsection{Comparisons with Other Systems\label{sec:compare}}

We have reported the performance of training GPT-2 with asynchronous checkpointing and restart in Figure~\ref{fig:checkpoint_perf}\textemdash the checkpointing-based approach spent only 23\% on actual training, while \system increases this percentage to \textbf{84\%}. 
In fact, as shown in Table~\ref{tab:simulation}, even for the preemption rate of 0.5, there are only 5.98 fatal failures that would require checkpointing/restart under \system. On the contrary, a checkpointing-based approach would need to restart the pipeline for every one of the 709.83 preemptions. Similarly, sample dropping significantly slows down the training when the preemption rate increases, as shown in Figure~\ref{fig:sample_dropping}.  

\MyPara{Varuna.} Varuna~\cite{varuna-2021} is a system developed concurrently with \system to enable training on spot instances. As with other existing techniques, Varuna provides resilience with checkpointing. 
We set up Varuna on the same spot cluster on AWS EC2 as we used in \S~\ref{sec:overall}. We ran Varuna with a $D \times P$ pipeline (\ie, the same as on-demand instances) because Varuna does not use redundancies and hence not need to over-provision resources. 

\begin{figure}[!ht]
    \centering
    \includegraphics[scale=.4]{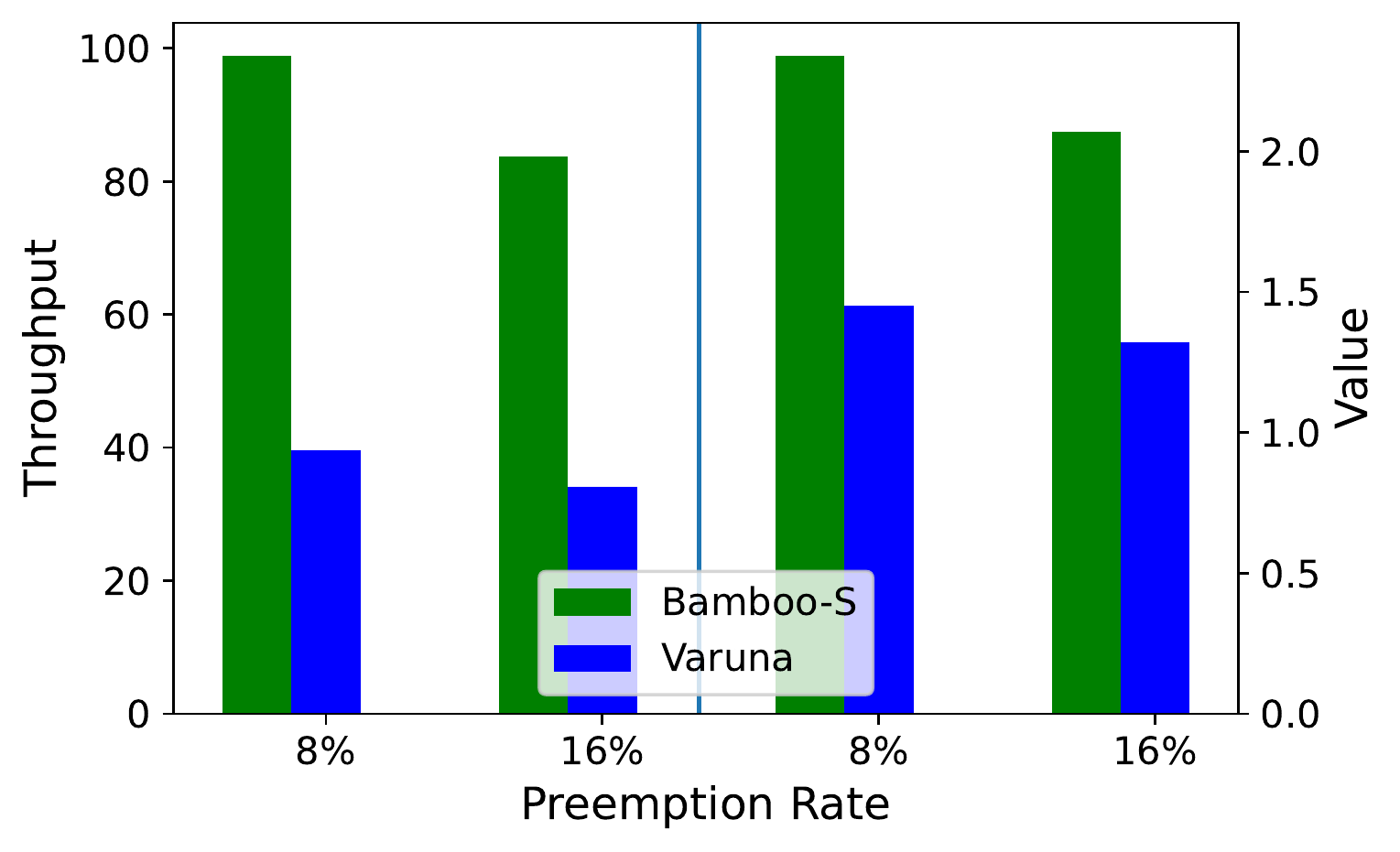}
    \caption{Throughput and value for Bamboo-S and Varuna running BERT at different preemption levels. Varuna hangs at the 33\% preemption rate. \label{fig:varuna}}
    \vspace{-.5em}
\end{figure}

We trained BERT on Varuna with the same configurations, including the same datasets, model architectures, float precision, preemption rates, and hyperparameters. Varuna hang under the 33\% preemption rate. For the 10\% and 16\% preemption rates, comparisons between Varuna and \system{}-S are reported in Figure~\ref{fig:varuna}. 
As shown, \system{}-S outperforms Varuna by \textbf{2.5$\times$} and \textbf{2.7$\times$} in throughput, respectively, under the 10\% and 16\% rates; and by \textbf{1.67$\times$} and \textbf{1.64$\times$}, in value, under these two rates. Note that value benefits are lower than throughput benefits due to Varuna's use of fewer instances.

\subsection{Microbenchmarks of Redundant Computation\label{sec:overhead}}
To fully understand the overhead introduced by RC, we compared time and memory among three versions of RC: eager-FRC-lazy-BRC (EFLB, \system's approach), eager-FRC-eager-BRC (EFEB), and lazy-FRC-lazy-BRC (LFLB), when training BERT and ResNet. Since the focus here is the RC overhead, we ran this experiment over on-demand instances. 

\begin{table}[h!]
\small
\centering
\begin{tabular}{lll}
\rowcolor[HTML]{EFEFEF}
 & BERT & ResNet \\
Lazy-FRC-Lazy-BRC & 7.01\% & 7.65\% \\
Eager-FRC-Lazy-BRC (\system) & 19.77\% & 9.51\% \\
Eager-FRC-Eager-BRC & 71.51\% & 64.24\%
\end{tabular}
\caption{Time overhead with different RC settings. \label{tab:time-overhead}}
\vspace{-1em}
\end{table}

Table~\ref{tab:time-overhead} reports RC's time overheads for the three RC settings. As expected, LFLB incurs the lowest per-iteration overhead because neither FRC nor BRC is performed with normal training iterations.  The $\thicksim$7\% overhead comes primarily from the extra code executed to prepare for a failover schedule.  However, the recovery time is much longer under LFLB than the other two settings (discussed shortly).  On the contrary, EFEB has the highest per-iteration overhead due to the eager execution of both FRC and BRC. The overhead incurred by EFLB, as used in \system, is slightly higher than LFLB but much lower than EFEB. This is because eager FRC does not incur extra communication overhead and much its computation overhead can be hidden by scheduling it into the pipeline bubble and overlapping it with FNC. 

Another interesting observation is the overhead for ResNet is lower than for BERT.  This is because ResNet's layer partitioning is much more imbalanced than that of BERT (which is a transformer model where most the middle layers are equivalent).  As a result, the bubble in ResNet's pipeline is much larger and hence it can accommodate a more significant fraction of FRC. 

Eager FRC incurs an overall $\thicksim$1.5$\times$ overhead in GPU memory (that is why \system recommends creating pipelines with 1.5$\times$ more nodes) while lazy FRC does not incur any memory overhead.


\begin{figure}[h!]
    \centering
    \begin{adjustbox}{max width=\linewidth}
    \begin{tabular}{cc}
        \includegraphics{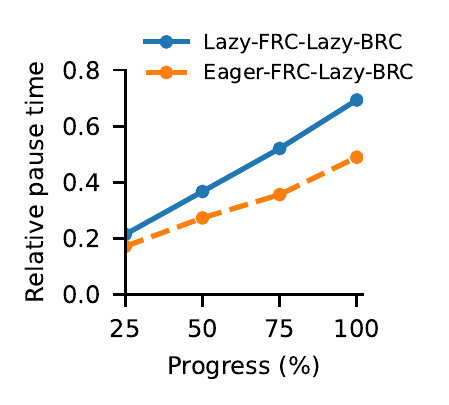} &
        \includegraphics{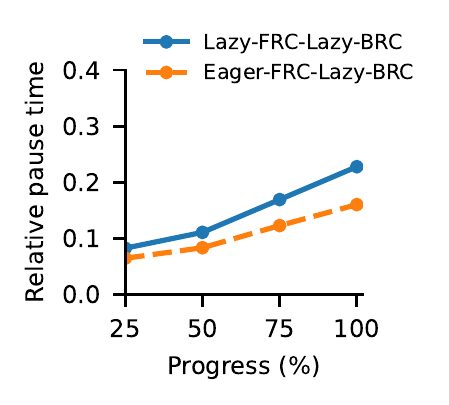} \\[-1em]
        (a) BERT & (b) ResNet
    \end{tabular}
    \end{adjustbox}
    \caption{Relative pause time for BERT and ResNet under different RC settings. \system runs into a pause when a pipeline stops training and waits for the shadow node to recover the lost state on the victim node.\label{fig:pause}}
    \vspace{-.5em}
\end{figure}

To understand the pause time under these different RC settings, Figure~\ref{fig:pause} shows the relative pause time (\ie, the actual pause time relative to the time of each training iteration without preemptions). As shown, lazy FRC reduces pause time by $\thicksim$35\% despite the slightly higher per-iteration overhead it introduces. In summary, eager-FRC-lazy-BRC strikes the right balance between overhead and pause time.

\subsection{Cross-Zone Communication \label{sec:cross}}

\begin{table}[h!]
\small
\vspace{-1em}
\centering
\begin{tabular}{cccc}
\rowcolor[HTML]{EFEFEF}
 Model & Config & Throughput & Total Transferred Bytes \\
BERT & Spread & 148.923 & 16.39 GiB \\
BERT & Cluster & 151.124 & 16.39 GiB \\\hline
VGG19 & Spread & 160.12 & 11.213 GiB \\
VGG19 & Cluster & 165.77 & 11.213 GiB \\
\end{tabular}
\vspace{-1em}
\caption{Comparison of throughput when running across availability zones compared
        to running within a single zone. \label{tab:cross-zone}}
\vspace{-1em}
\end{table}

Because \system allocates workers across availability zones to minimize the
probability of reconfigurations, we measured 
We ran \system in two configurations: (1) with nodes distributed across all zones (\ie, Spread)
and (2) in a single availability zone with AWS' ``Placement Group'' option set to
``Cluster'' (\ie, Cluster), and measured their performance differences. As reported in Table~\ref{tab:cross-zone}, the differences between these two configurations are quite low (\ie, usually less than 5\%). This demonstrates \system's choice of assigning nodes from different availability zones as consecutive nodes in each pipeline has little impact on training performance.

\subsection{\system for Pure Data Parallelism \label{sec:puredp}}

We ran two relatively small models such as VGG and ResNet using pure data parallelism with 8 workers (\ie, we partition the data but not the model). For \system, we similarly over-provisioned 1.5$\times$ additional workers. We implemented another baseline \emph{Checkpoint}, which periodically checkpoints model state for each worker and restarts the worker on another node when its original node is preemption. We used the same global batch size for these models as reported in \S\ref{sec:evaluation}. The comparisons between \system, \emph{Checkpoint}, and on-demand training are shown  in Table~\ref{tab:dataparallel}. 

Note that our implementation of \emph{Checkpoint} assumes that there is always a standby node that is ready to join and load the checkpoint (which is a unrealistic over-approximation of the allocation model on any spot market); as such, the training cost remains unchanged and its throughput is reduced as the preemption rate increases.

\begin{table}[h!]
\begin{adjustbox}{max width=\linewidth}
  \centering
  \begin{tabular}{ll|rrr}
  \small
    Model & System & Throughput & Cost (\$/hr) & Value \\
  \multirow{2}{*}{ResNet} & Demand & 24.51 & 24.48 & 1.01  \\
                          & Checkpoint & [12.26, 8.42, 5.03] & [7.34, 7.34, 7.34] & [1.67, 1.15, 0.68] \\
                          & \system  &  [\textbf{21.22}, 18.31, 12.31] & [\textbf{10.56}, 10.09, 9.18]  &  [\textbf{2.01}, 1.84, 1.34] \\\hline
  \multirow{2}{*}{VGG} & Demand  & 144.28 & 24.48 & 5.89 \\
                       & Checkpoint & [83.21, 67.21, 45.31] & [7.34, 7.34, 7.34] & [11.33, 9.15, 6.17]\\
                       & \system    & [\textbf{125.59}, 96.51, 73.73]  & [\textbf{10.56}, 10.09, 9.18] & [\textbf{11.89}, 9.56, 8.03] \\\hline \\[-1em]
\end{tabular}
\vspace{-1.5em}
\end{adjustbox}
\caption{Comparison between pure data-parallel training over on-demand instances, a checkpoint-based approach on spot instances, \system on spot instances. For \emph{Checkpoint} and \system, we trained each model three times, and their results are explicitly listed in the form of [$a$, $b$, $c$] for the 10\% (average), 16\%, and 33\% preemption rates, respectively.\label{tab:dataparallel}}
\vspace{-.5em}
\end{table}

As shown, \system outperforms \emph{Checkpoint} by \textbf{1.64$\times$} and \textbf{1.22$\times$} in throughput and value. Both \emph{Checkpoint} and \system deliver a higher value than on-demand training (by \textbf{2$\times$} and \textbf{1.79$\times$}). 

We make two observations on these numbers. First, \system incurs a higher cost than \emph{Checkpoint} due to resource over-provisioning. However, as discussed above, \emph{Checkpoint} assumes the availability of standby nodes. In practice, guaranteeing such availability requires over-provisioning as well, but we did not take this into account when calculating costs (because it is hard to know exactly how many nodes we should over-provision). Hence, the cost and value reported for \emph{Checkpoint} are the \emph{lowerbound} and \emph{upperbound} of those that can be achieved by any practical implementation of a checkpoint-based approach. 

Second, \emph{Checkpoint} works much better for pure data parallelism than for pipeline parallelism (as discussed in \S\ref{sec:motivation}). This is because recovering from a checkpoint in pure data-parallel training is much easier than pipeline-parallel training where a pipeline reconfiguration process is needed for each restart.

\section{Related Work}
\label{sec:related-work}

\MyPara{Parallel Training.}
Data parallelism~\cite{flexflow-mlsys18,dean-nips12,imagenet-cnn-cacm17, all-reduce-16,geeps-eurosys16,poseidon-atc17,li-osdi14,poseidon-atc17} is the most common parallelism model that partitions the dataset and trains on each partition. The learned weights are synchronized via either an all-reduce approach~\cite{all-reduce-16} or parameter servers~\cite{li-osdi14,adam-osdi14}. Model parallelism~\cite{dean-nips12,parallelize-cnn14,elastic-mlsys20,nips/2018/shazeer, corr/abs-1909-08053} partitions the operators in a DNN model
across multiple GPU devices, with each worker evaluating and
performing updates for only a subset of the model’s parameters
for all inputs. Recently, pipeline parallelism~\cite{gpipe-18,pipedream-sosp19,pipemare-mlsys19,dorylus-osdi21} has been proposed to train large models by partitioning layers across workers and uses microbatches to saturate the pipeline. Popular DL training libraries such as DeepSpeed~\cite{deepspeed-sc20} and Megatron~\cite{narayanan-sc21} support 3D parallelism, which combines data parallelism, model parallelism, and pipeline parallelism to train models at extremely large scale with improved compute and memory efficiency. Furthermore, DeepSpeed offers ZeRO-style data parallelism~\cite{rajbhandari2020zero}, which partitions model states across GPUs and uses communication collectives to gather individual parameters when needed. 


\MyPara{Elastic Training.} Distributed training experiences frequent resource changes. There are a number of systems~\cite{elastic-mlsys20,proteus-eurosys17,pytorch-elastic,huang-sigmod15,litz-atc18,elastic-nsdi21}  built to provide elasticity for training over  changing resources. TorchElastic~\cite{pytorch-elastic} is a PyTorch~\cite{pytorch-neurips19}-based tool that can dynamically kill or add data-parallel workers. Huang \emph{et al.}~\cite{huang-sigmod15} considers elasticity for
declarative ML on MapReduce, which does not work for modern deep learning workloads. Litz~\cite{litz-atc18} is a system that provides elasticity in the context of CPU-based machine learning using the parameter servers. Or \emph{at al.}~\cite{elastic-mlsys20} presents an autoscaling system built on top of TensorFlow~\cite{tensorflow-osdi16} and Horovod~\cite{horovod}, which dynamically adapts the batch size and reuses existing processes.

\MyPara{Exploiting Spot Instances.} 
Proteus~\cite{proteus-eurosys17} exploits dynamic pricing on public
clouds in order to lower costs for machine learning workloads through elasticity. Since Proteus does not explicitly consider modern deep learning workloads, Proteus simply reprocesses the input of a preempted node with another node. Varuna~\cite{varuna-2021} is a system built concurrently with \system for distributed training over spot instances. However, Varuna focuses on elasticity, not quick recovery from preemptions. \system, on the contrary, is designed specifically to deal with frequent preemptions. 

There exists a body of work on enabling low latency and/or SLO guarantees when using preemptible spot instances. Tributary~\cite{tributary-atc18} is an elastic control system that exploits preemptible resources to
reduce cost with SLO guarantees. Kingfisher~\cite{kingfisher} proposes a cost-aware resource acquisition
scheme that uses integer linear programming to
determine a service’s resource footprint among a heterogeneous set of non-preemptible instances with fixed
prices. Flint~\cite{flint-eurosys16} is a system that runs batch-based data-intensive jobs on transient servers. SpotCheck~\cite{spotcheck-eurosys15} selects
spot markets to acquire instances in while always bidding at a configurable multiple of the spot instance’s corresponding on-demand price.  BOSS~\cite{boss-infocom16} hosts key-value stores on spot instances
by exploiting price differences across pools in different
data-centers. ExoSphere~\cite{exosphere-sigmetric17} is a virtual cluster framework for
spot instances. These systems are all orthogonal to \system that is built specifically for deep learning training. 

\MyPara{GPU Scheduling.} There is also a large body of work on GPU scheduling~\cite{nexus-sosp19,antman-osdi20,zhang-mobihoc20,optimus-eurosys18,themis-nsdi20,tiresias-nsdi19,narayanan-osdi20,narayanan-sc21,pretzel-osdi18,slaq-socc17} for ML workloads. These techniques are orthogonal to \system\textemdash they all focus on efficiency and throughput while \system aims to perform redundant computation at a low cost. 
\section{Conclusion}
\label{sec:conclusion}

\system{} is the first distributed system that uses redundant computation to provide resilience and fast recovery for training large DNN models on preemptible instances. An evaluation with 6 representative models shows that \system{} provides a much higher value than (1) training on on-demand instances and (2) training with checkpointing/restart on spot instances.

\printbibliography

\end{document}